\documentclass[acmsmall,screen]{acmart}

\makeatletter
\AtBeginDocument{%
  \fancypagestyle{firstpagestyle}{%
    \fancyhf{}%
    \fancyhead[L]{\ACM@linecountL}%
  }%

  \fancypagestyle{standardpagestyle}{%
    \fancyhf{}%
    \fancyhead[L]{\ACM@linecountL}%
    \fancyhead[R]{\thepage}%
  }%

  \pagestyle{standardpagestyle}%
}
\makeatother
\AtBeginDocument{%
  }

\renewcommand\footnotetextcopyrightpermission[1]{}

\usepackage[utf8]{inputenc}     
\usepackage[T1]{fontenc}
\usepackage{bigstrut}           
\usepackage{amsmath}            

\usepackage{wasysym}
\usepackage{hhline}  
\usepackage{subfig} 

\usepackage{varioref}

\usepackage{xurl}
\usepackage{xcolor}
\usepackage{colortbl}
\usepackage{listings}
\usepackage{booktabs}
\input{code-style-d} 
\hypersetup{bookmarksopen=true,bookmarksopenlevel=0,pdfpagelayout=SinglePage}

\usepackage{longtable}
\usepackage{multirow}
\usepackage{sidecap}            
\usepackage{tabularx}
\usepackage{ragged2e}
\usepackage{rotating}

\usepackage{fancyhdr}
\usepackage{graphicx} 
\usepackage{caption}
\usepackage{wrapfig}
\usepackage{pdfpages}
\usepackage{enumitem}
\usepackage{ifthen}
\usepackage{rotating}
\usepackage{listings}
\usepackage{calc}                  

\usepackage{float}
\usepackage{csvsimple-l3}     
\usepackage{etoolbox}         
\usepackage[acronym,nohypertypes=acronym]{glossaries}
\usepackage{appendix}
\usepackage{etoc}             
\usepackage{lipsum}           
\usepackage[symbol*]{footmisc} 
\usepackage{color}

\usepackage{xcolor}

\definecolor{mauve}{rgb}{0.58, 0, 0.82}

\usepackage{pifont}
\usepackage{xspace}
\usepackage{tikz}
\newcommand{\cachetikzfigure}[2]{}
\newcommand{\sys}{ForTress\xspace}

\usepackage{xcolor}
\usepackage{listings}
\usepackage{caption}
\usepackage{xspace}
\lstdefinelanguage{CaseSnippet}{
  morecomment=[l]{//},
  morecomment=[l]{\#},
  sensitive=true
}

\lstdefinestyle{casecode}{
  language=CaseSnippet,
  basicstyle=\ttfamily\footnotesize,
  keywordstyle=\color{blue!70!black},
  morekeywords={ANINT,np,round,trunc,abs,copysign,if}, 
  commentstyle=\color{green!45!black},
  columns=fullflexible,
  keepspaces=true,
  showstringspaces=false,
  breaklines=true,
  breakatwhitespace=false,
  linewidth=\linewidth,
  frame=single,
  numbers=none,
  framerule=0.6pt,
  rulecolor=\color{black!65},
  framesep=6pt,
  xleftmargin=0pt,
  xrightmargin=0pt,
  framexleftmargin=0pt,
  framexrightmargin=0pt,
  framextopmargin=3pt,
  framexbottommargin=3pt,
  aboveskip=0pt,
  belowskip=0pt,
  lineskip=-1pt
}

\definecolor{darkgreen}{rgb}{0.0,0.5,0.0}
\definecolor{darkblue}{rgb}{0.0, 0.0, 0.55}

\newcommand{\eg}{\textit{e.g.}}

\begin{document}

\title{\sys: Bringing Assurance and Determinism for LLM-Driven Fortran-to-Python Translation}

\author{Qinrun Dai}
\email{Qinrun.Dai@colorado.edu}
\affiliation{%
  \institution{University of Colorado Boulder}
  \city{Boulder}
  \state{Colorado}
  \country{USA}
}

\author{Ruitong Liu}
\email{ruli3957@colorado.edu}
\affiliation{%
  \institution{University of Colorado Boulder}
  \city{Boulder}
  \state{Colorado}
  \country{USA}
}

\author{Chien-Wei Huang}
\email{Jeremy.Huang-1@colorado.edu}
\affiliation{%
  \institution{University of Colorado Boulder}
  \city{Boulder}
  \state{Colorado}
  \country{USA}
}

\author{Guanhong Tao}
\email{guanhong.tao@utah.edu}
\affiliation{%
  \institution{University of Utah}
  \city{Salt Lake City}
  \state{Utah}
  \country{USA}
}

\author{Guannan Wei}
\email{guannan.wei@tufts.edu}
\affiliation{%
  \institution{Tufts University}
  \city{Medford}
  \state{Massachusetts}
  \country{USA}
}

\author{Yueqi Chen}
\email{yueqi.chen@colorado.edu}
\affiliation{%
  \institution{University of Colorado Boulder}
  \city{Boulder}
  \state{Colorado}
  \country{USA}
}



\begin{abstract}
Modernizing mature scientific software from Fortran to Python can improve accessibility and integration with modern computing tools, but manual rewriting is expensive.
Large language models (LLMs) make translation more practical, yet generated code can pass regression tests while violating source-language semantics, and contextual corrections may not persist across subsequent translations.
We present \sys, which solidifies recurring LLM translation decisions into guarded executable rules for deterministic reuse and provides semantic assurance for a non-trivial subset of these rules.
Its LLM-assisted auditing process expresses universally quantified preservation obligations over mechanized Fortran and Python semantics in Lean.
Human experts review semantic fidelity and correspondence with production rules, and Lean's kernel checks proofs and refutations under explicit assumptions.
These audit records expose the conditions either supporting correctness or identifying failure cases.
We evaluate \sys on 17 benchmarks from 15 scientific software spanning diverse domains.
The translated programs achieve bitwise-identical with native Fortran on all evaluated tests under the recorded execution and initialization configurations.
Against these software, \sys accumulates 193 executable rules and 188 audit records, of which 121 have checked soundness proofs, including 31 established after counterexample-guided refinement.
Cross-software reuse exceeds 90\% for 14 of the 15 software.
Even with additional assurance and determinism, the LLM usage cost is 44.2\% lower than direct LLM translation.

\end{abstract}

\ccsdesc[500]{Software and its engineering~Source code generation}
\ccsdesc[300]{Software and its engineering~Software testing and debugging}


\keywords{Fortran-to-Python translation, scientific software modernization, reusable translation rules, semantic assurance, LLM-assisted program translation, Lean}


\maketitle

\section{Introduction}

Large scientific software systems are increasingly being modernized from legacy languages such as Fortran into Python-based ecosystems, as exemplified by Veros~\cite{hafner2018veros} and PyCAM5~\cite{pycam5status}. 
Such migrations promise improved maintainability, accessibility, accelerator support, and integration with modern AI tools~\cite{hafner2018veros,hafner2021veros,meunier2025differentiableveros}. 
However, manually rewriting mature scientific codes is prohibitively expensive.
The U.S. Department of Energy's Exascale Computing Project invested up to \$1.8 billion over seven years to develop and enhance 25 scientific software~\cite{koppes2024ecpnasa,llnl2025ecp}. 
Recent Large Language Model (LLM)-driven coding systems make source-to-source translation possible at a scale that was previously impractical~\cite{ibrahimzada2025alphatrans,lahlou2026scientifictranslation,koldunov2026fesom2}.

However, LLM-driven translation faces two key bottlenecks in practice. 
First, an LLM-generated translation may be syntactically valid and pass regression tests while still failing to preserve the original program semantics.
Execution-based dynamic validation remains indispensable, but it establishes agreement only for the finite inputs actually tested. 
Second, even after a translation error is identified and corrected, documenting the correct translation in the context or as a “skill” does not ensure its consistent reuse in subsequent translations.
The LLM could still overlook the correction or generate another plausible but semantically incorrect translation.


In this work, we propose a two-part approach to address these bottlenecks. 
First, we solidify recurring Fortran-to-Python transformations observed during LLM-driven translation and debugging into executable heuristic rules, allowing subsequent deterministic reuse.
Second, for a nontrivial subset of these rules, we use Lean to check proofs that the rules preserve semantics under explicitly stated applicability conditions and modeling assumptions.
More specifically, the LLM interacts with Lean by proposing proofs covering all source instances and inputs satisfying these conditions, which Lean then checks.
If these proof attempts do not succeed, the LLM searches for counterexamples that expose failure conditions and guide rule and proof refinement, providing a basis for reuse that black-box LLM generation alone does not offer.


The formal semantic definitions and theorem statements
are initially generated with LLM assistance and are not trusted as generated.
For modeled construct, human experts check that the definitions reflect the relevant Fortran and Python behavior, including types, operations, and state changes, under explicit assumptions. 
They also check that each preservation claim accurately describes the corresponding executable rule's source pattern, applicability conditions, and emitted Python code.
As such, an incorrect rule is still faithfully described, allowing a counterexample to refute its universal preservation claim and identify a source instance and input for which preservation fails.

We implemented the proposed approach as \sys.
\sys is a continuously evolving system that accumulates and refines LLM-generated components using feedback from translating and evaluating scientific software.
To date, we have evaluated \sys on 17 benchmarks from 15 software spanning diverse domains, using Claude Opus~5.5, the strongest model available to us at the time of the experiments and submission.
\sys translated 105,631 lines of Fortran code from 1,015  subprograms reached by the benchmarks.
The translated Python programs produced outputs bitwise-identical to those of native Fortran on all evaluated benchmark tests under the instrumented execution and initialization configurations.

We accumulated 193 executable rules, 42 Lean modules, and 188 audit records across these software.
Of these records, 121 have checked soundness proofs, including 41 under additional applicability premises and 31 established after counterexample-guided refinement. 
There are another 25 records requiring explicit model extensions and external contracts, and 42 records remaining to be further refined and checked against more software instances.

In leave-one-software-out reuse measurements, 14 of the 15 software exceed 90\% reuse, and six reach 100\%.
Even with additional assurance and determinism, \sys's LLM usage cost is \$960, compared with \$1,721 for the direct LLM baseline, a reduction of 44.2\%.
The developers of CLUBB~\cite{software-clubb}, one of our evaluated software, have already integrated part of the translated Python code into their codebase.
Though \sys is developed and evaluated for Fortran-to-Python translation, the approach is general and can be potentially applied to other language pairs.

We make the following technical contributions:
\begin{itemize}
    \item  We introduce reusable, guarded executable translation rules that capture recurring LLM translation decisions and enable deterministic reuse, amortizing the cost of LLM-assisted translation.
    
    \item We provide assurance for executable translation rules by linking them to mechanized Fortran and Python semantics through explicit preservation obligations. Human review checks semantic fidelity and rule--obligation correspondence, while Lean's kernel checks proofs and refutations under stated assumptions.
    
    \item We implement \sys and showcase its translation effectiveness, rule assurance, and cross-software reuse on 17 benchmarks from 15 scientific software systems. 
    \sys is released as part of RecastEngine~\cite{recastengine2026}.
\end{itemize}

We organize the rest of this paper as follows.
Section~\ref{sec:motivation} motivates our work.
We describe the \sys approach in Section~\ref{sec:approach} and its LLM-generated components in Section~\ref{sec:analysis}.
Section~\ref{sec:evaluation} presents the evaluation results.
Section~\ref{sec:threats} discusses threats to validity, followed by related work in Section~\ref{sec:related}.
Finally, Section~\ref{sec:conclusion} concludes this paper.

\begin{figure}[t]
\centering
\begingroup

\newsavebox{\caseCodeBox}
\begin{lrbox}{\caseCodeBox}
\begin{minipage}[t]{0.60\linewidth}
\centering
\normalsize\textbf{Translation example}

\vspace{5pt}
\begin{lstlisting}[style=casecode,
  basicstyle=\normalsize\ttfamily,
  frame=tb, framesep=3pt,
  rulecolor=\color{black}, framerule=\heavyrulewidth,
  xleftmargin=0pt, xrightmargin=0pt,
  framextopmargin=0pt, framexbottommargin=0pt,
  aboveskip=2.38pt, belowskip=0pt, lineskip=-1.5pt, emptylines=0]
// Fortran source
r1 = ANINT(x)

// Incorrect translation
r1 = np.round(x)

// Repaired translation
t = np.trunc(x)
r1 = (t + math.copysign(1.0, x)
      if abs(x - t) >= 0.5 else t)
\end{lstlisting}
\vspace*{0pt}
\end{minipage}
\end{lrbox}

\normalsize
\setlength{\tabcolsep}{0pt}
\newlength{\caseTableColWidth}
\settowidth{\caseTableColWidth}{\textbf{\texttt{np.copysign}}}
\renewcommand{\arraystretch}{1.05}
\setlength{\extrarowheight}{3.2pt}

\newsavebox{\caseTableBox}
\begin{lrbox}{\caseTableBox}
\begin{tabular}{@{}*{5}{>{\centering\arraybackslash}m{\caseTableColWidth}}@{}}
    \toprule
    \textbf{Input \texttt{x}} &
    \textbf{\texttt{ANINT}} &
    \textbf{\texttt{np.round}} &
    \textbf{\texttt{np.trunc}} &
    \textbf{Repaired \texttt{r1}} \\
    \midrule
    \texttt{2.25} & \texttt{2.0}  & \texttt{2.0}  & \texttt{2.0}  & \texttt{2.0} \\
    \texttt{1.5}  & \texttt{2.0}  & \texttt{2.0}  & \texttt{1.0}  & \texttt{2.0} \\
    \texttt{2.5}  & \texttt{3.0}  & \texttt{2.0}  & \texttt{2.0}  & \texttt{3.0} \\
    \texttt{-2.5} & \texttt{-3.0} & \texttt{-2.0} & \texttt{-2.0} & \texttt{-3.0} \\
    \bottomrule
\end{tabular}
\end{lrbox}

\newsavebox{\caseCombinedBox}
\begin{lrbox}{\caseCombinedBox}
\begin{minipage}[t]{\wd\caseCodeBox}
\usebox{\caseCodeBox}
\end{minipage}
\hspace{10pt}
\begin{minipage}[t]{\wd\caseTableBox}
\centering
\normalsize\textbf{Rounding behavior}

\vspace{5pt}

\usebox{\caseTableBox}

\vspace{6pt}

\begin{minipage}{\wd\caseTableBox}
\raggedright
\end{minipage}
\end{minipage}
\end{lrbox}

\resizebox{\linewidth}{!}{\usebox{\caseCombinedBox}}

\endgroup

\caption{Rounding mismatch in direct LLM translation. Left: a Fortran
\texttt{ANINT} call, the incorrect \texttt{np.round} translation, and a
repaired scalar rounding core. Right: representative outputs showing that
agreement on ordinary inputs does not ensure correct halfway rounding.}
\Description{Fortran ANINT and NumPy round agree at 2.25 and 1.5 but differ at
positive and negative 2.5. A repaired scalar core truncates toward zero and
adjusts away from zero when the residual magnitude is at least one half.}
\label{fig:case1}
\end{figure}

\section{Motivating Example}
\label{sec:motivation}

Consider the translation of the Fortran assignment statement in Figure~\ref{fig:case1}.
According to the Fortran 2018 standard (ISO/IEC 1539-1:2018)~\cite{NintF2018}, the intrinsic function \texttt{ANINT} returns a whole-number real value and rounds halfway
cases away from zero.
Python's \texttt{np.round}, in contrast, rounds ties to the nearest even value~\cite{NintNumPyRound}. Thus, the seemingly natural translation from \texttt{ANINT} to \texttt{np.round} is semantically incorrect.
The error is difficult to detect through testing because the two implementations agree on many inputs.
For example, both return \texttt{2.0} for \texttt{x = 2.25}, and even agree on the halfway input \texttt{x = 1.5}.
However, they diverge on discriminating ties.
At \texttt{x = 2.5}, Fortran returns \texttt{3.0} whereas NumPy returns \texttt{2.0}, and at \texttt{x = -2.5}, they return \texttt{-3.0} and \texttt{-2.0}, respectively.

Such divergence matters because it directly affects computational correctness.
In Quantum ESPRESSO~\cite{giannozzi2009qe,giannozzi2017qe}, a widely-used software for electronic-structure calculations and materials modeling, \texttt{ANINT} is used in its dipole routine when selecting minimum-image coordinates, and the rounding divergence mistakenly changes all three dipole components from $-4$ to $4$.
Therefore, semantic assurance are required to preserve scientific fidelity during translation.

Identifying and correcting one divergence instance is insufficient.
The correction must be reliably reused across all sites.
Prior LLM-driven translation workflows use reusable ``skills'' to guide the process and persistent context files to carry translation guidance and corrections across sessions~\cite{koldunov2026fesom2,lahlou2026scientifictranslation}.
However, neither approach enforces consistent application of a correction, because context can be lost over long-horizon translation and stochastic token generation provides no guarantee.

Using more capable frontier models reduces the likelihood of such mistranslations to some extent, but does not fundamentally address the problem.
We experimented two models, Opus~4.6 and ~5.5 on 50 code snippets containing \texttt{ANINT} and its sibling intrinsics \texttt{NINT}, repeating each translation six times.
The average number of mistranslations decreases only from $46.5$ to $34.5$, even after unlimited LLM attempts.
Moreover, Opus~5.5 introduces another seemingly plausible translation, \texttt{math.floor(x+0.5)}, which also diverges from the original Fortran semantics when \texttt{x} is very close to, but smaller than, $0.5$.
Even if future models can reliably handle this particular translation, stronger model capability does not generalize into a guarantee against other latent semantic mismatches, nor does it ensure persistent correctness throughout a long-horizon translation.

In this work, while retaining LLMs as part of the translation process, we propose to solidify translations into executable rules for reliable reuse.
For each executable rule, we construct a soundness obligation relating the source operation to the target operation implemented by the rule.
For these obligations, Lean checks the preservation proofs and counterexample-based refutation proofs constructed by the LLM.
Both outcomes provide assurance about the rule's semantic behavior.
A proof establishes preservation for every modeled instance satisfying its premises, while a refutation identifies a modeled counterexample to the preservation claim which guides further repair.
This evidence makes the conditions supporting correctness or identifying failure cases explicit, providing a basis that black-box LLM generation alone does not offer.

\begin{figure}[t]
\centering
\begingroup

\normalsize
\setlength{\tabcolsep}{0pt}

\newlength{\ruleRefColWidth}
\settowidth{\ruleRefColWidth}{\textbf{\texttt{np.copysign}}}

\newlength{\ruleRefRightWidth}
\setlength{\ruleRefRightWidth}{\dimexpr5\ruleRefColWidth\relax}

\newsavebox{\ruleCodeBox}
\begin{lrbox}{\ruleCodeBox}
\begin{minipage}[t]{0.60\linewidth}

\centering
\normalsize\textbf{Executable rules}

\vspace{5pt}

\begin{lstlisting}[style=casecode,
  basicstyle=\normalsize\ttfamily,
  frame=tb, framesep=3pt,
  rulecolor=\color{black}, framerule=\heavyrulewidth,
  xleftmargin=0pt, xrightmargin=0pt,
  framextopmargin=0pt, framexbottommargin=0pt,
  aboveskip=6pt, belowskip=0pt, lineskip=-1.5pt]
# Intrinsic translation
def translate_intrinsic(name, args):
    rule = scalar_table[name]
    return f"{rule}({args[0]})"

scalar_table = {
    "anint": "_f_anint"
}

# Incorrect executable rule
def _f_anint(x):
    return np.round(x)

# Repaired executable rule
def _f_anint(x):
    t = np.trunc(x)
    return (t + math.copysign(1.0, x)
        if abs(x - t) >= 0.5 else t)
\end{lstlisting}
\vspace{-3pt}

\end{minipage}
\end{lrbox}

\newsavebox{\ruleLeanBox}
\begin{lrbox}{\ruleLeanBox}
\begin{minipage}[t]{\ruleRefRightWidth}

\centering
\normalsize\textbf{Lean assurance}

\vspace{5pt}

\begin{lstlisting}[style=styleD,language=LeanD,numbers=none,
  xleftmargin=0pt,xrightmargin=0pt,
  rulecolor=\color{black},framerule=\heavyrulewidth]
def FNP_R0064_obligation : Prop :=
  (*@$\forall$@*) o (*@$\Gamma$@*), Call1Sound o (*@$\Gamma$@*) .anint .np_round
theorem FNP_R0064_refuted :
  ¬ FNP_R0064_obligation
\end{lstlisting}

\vspace{3pt}

\begin{lstlisting}[style=styleD,language=LeanD,numbers=none,
  xleftmargin=0pt,xrightmargin=0pt,
  rulecolor=\color{black},framerule=\heavyrulewidth]
noncomputable def anintRepairR (x : (*@$\mathbb{R}$@*)) : (*@$\mathbb{R}$@*) :=
  let t : (*@$\mathbb{R}$@*) := truncR x
  if (1 / 2 : (*@$\mathbb{R}$@*)) (*@$\leq$@*) |x - t| then
    t + (if x < 0 then -1 else 1)
  else t
\end{lstlisting}

\vspace{3pt}

\begin{lstlisting}[style=styleD,language=LeanD,numbers=none,
  xleftmargin=0pt,xrightmargin=0pt,
  rulecolor=\color{black},framerule=\heavyrulewidth,
  framexbottommargin=3.4pt]
def FNP_R0064_fixed_obligation : Prop :=
  (*@$\forall$@*) x : (*@$\mathbb{R}$@*), anintRepairR x = (nintR x : (*@$\mathbb{R}$@*))
theorem FNP_R0064_fixed_sound :
  FNP_R0064_fixed_obligation :=
  anintRepairR_eq_nintR
\end{lstlisting}

\end{minipage}
\end{lrbox}

\newsavebox{\ruleCombinedBox}
\begin{lrbox}{\ruleCombinedBox}
\begin{minipage}[t]{\wd\ruleCodeBox}
\usebox{\ruleCodeBox}
\end{minipage}
\hspace{10pt}
\begin{minipage}[t]{\wd\ruleLeanBox}
\usebox{\ruleLeanBox}
\end{minipage}
\end{lrbox}

\resizebox{\linewidth}{!}{\usebox{\ruleCombinedBox}}

\endgroup

\caption{Reusable translation rules and their Lean soundness audit for
\texttt{ANINT}. Left: deterministic intrinsic dispatch and the incorrect
and repaired executable rules. Right: Lean checks LLM-generated proofs refuting the original rule's soundness obligation and establishing that the repaired rounding core is equivalent to the modeled Fortran semantics.}
\label{fig:anint-rule}
\end{figure}

Figure~\ref{fig:anint-rule} illustrates how such an executable translation rule is represented and applied for our \texttt{ANINT} example.
It is essentially an Python logic that dispatches a source-language construct to its corresponding target-language implementation.
In the figure, \texttt{translate\_intrinsic()} looks up the corresponding rule in \texttt{scalar\_table} and emits a call to that rule with the translated argument.
For \texttt{ANINT}, the table maps the intrinsic to the Python helper \texttt{\_f\_anint}, so every occurrence of \texttt{ANINT(x)} is deterministically translated into \texttt{\_f\_anint(x)}.
The incorrect rule simply returns \texttt{np.round(x)}, whereas the repaired rule truncates the input and adjusts the result away from zero whenever the fractional magnitude reaches one half.

We next formulate the semantic correctness of the original rule as
\texttt{FNP\_R0064\_obligation}.
As shown in Figure~\ref{fig:anint-rule}, the obligation requires
\texttt{Call1Sound o $\Gamma$ .anint .np\_round} for all semantic oracles
\texttt{o} and source environments \texttt{$\Gamma$}.
\texttt{Call1Sound} captures compositional soundness for a single-argument
operation.
If a source argument and its translated target expression are
already semantically related by \texttt{SoundScalar}, then applying the
source intrinsic and the target helper must preserve that relation.
Thus, proving \texttt{FNP\_R0064\_obligation} would establish that the
translation from \texttt{ANINT} to \texttt{np.round} is sound for every
scalar expression covered by the semantic model.

For this rule, however, the LLM constructs a proof of the negated obligation, recorded as \texttt{FNP\_R0064\_refuted} and checked by Lean.
This formalizes the semantic mismatch in
Figure~\ref{fig:case1}.
Fortran \texttt{ANINT} uses the source rounding semantics \texttt{nintR}, whereas \texttt{np.round} uses \texttt{roundHalfEvenR}.
The Fortran and Python semantics used in these obligations are defined independently, based on their respective specifications and validated against compiler and runtime behavior.
At the exactly representable input $5/2$, \texttt{nintR(5/2)} evaluates to $3$, while \texttt{roundHalfEvenR(5/2)} evaluates to $2$. 
Instantiating the universal \texttt{Call1Sound} obligation with this argument therefore yields a contradiction.
Consequently, \texttt{FNP\_R0064\_refuted} is not merely a failed attempt to prove the rule, but a machine-checked proof that the rule is unsound.

This refutation guides the repair shown in Figure~\ref{fig:anint-rule}.
We model the scalar rounding core of the repaired implementation by \texttt{anintRepairR}.
For an input $x$, it first computes \texttt{t = truncR x}.
If $|x-t| \geq 1/2$, it adjusts \texttt{t} by one away from zero.
Otherwise, it returns \texttt{t} unchanged.
This is the exact-real counterpart of the repaired executable Python rule.

The repaired core is specified by \texttt{FNP\_R0064\_fixed\_obligation}.
In this proposition, \texttt{nintR} is the independently defined source-side rounding
semantics used to model \texttt{ANINT}.
The theorem \texttt{FNP\_R0064\_fixed\_sound} discharges this obligation
directly using the previously established lemma \texttt{anintRepairR\_eq\_nintR}.
Hence, the repaired rounding core is equivalent to the source rounding semantics for every real input in the model, not just for the example in Figure~\ref{fig:case1}.

\section{\sys Approach}
\label{sec:approach}

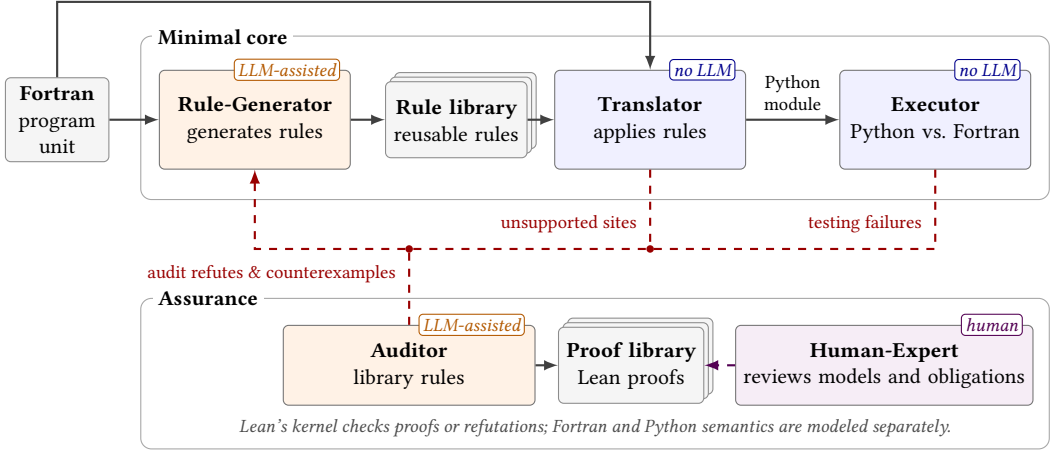
\begin{figure}[t]
  \centering
  \resizebox{\linewidth}{!}{%
\cachetikzfigure{approach}{figs/fig_approach.tex}
\begin{tikzpicture}[x=1cm,y=1cm,font=\footnotesize,
  box/.style={draw=black!55,rounded corners=2pt,align=center,
    text width=#1,minimum height=1.3cm,inner sep=3pt},
  box/.default=2.3cm,
  llm/.style={fill=orange!10},
  nollm/.style={fill=blue!6},
  person/.style={fill=violet!7},
  artifact/.style={draw=black!45,rounded corners=2pt,align=center,
    text width=#1,minimum height=1.0cm,inner sep=3pt,fill=black!4},
  tag/.style={font=\scriptsize\itshape,fill=white,draw=#1,text=#1,
    rounded corners=1.5pt,inner sep=1.5pt,anchor=east},
  group/.style={draw=black!35,rounded corners=4pt},
  legend/.style={font=\footnotesize\bfseries,fill=white,inner sep=2pt,anchor=west},
  flow/.style={-latex,thick,draw=black!75},
  feedback/.style={thick,dashed,draw=red!60!black},
  review/.style={-latex,thick,dashed,draw=violet!65!black},
  fblabel/.style={font=\scriptsize,text=red!60!black,anchor=east,inner sep=2pt}]

  \draw[group] (1.8,-1.05) rectangle (13.75,1.1);
  \node[legend] at (1.95,1.1) {Minimal core};
  \node[artifact=1.15cm] (source) at (0.7,0) {\textbf{Fortran}\\program unit};
  \node[box,llm] (generator) at (3.3,0) {\textbf{Rule-Generator}\\generates rules};
  \node[artifact=1.65cm] at (6.07,0.12) {};
  \node[artifact=1.65cm] at (6.01,0.06) {};
  \node[artifact=1.65cm] (library) at (5.95,0) {\textbf{Rule library}\\reusable rules};
  \node[box,nollm] (translator) at (8.5,0) {\textbf{Translator}\\applies rules};
  \node[box,nollm] (executor) at (12.25,0) {\textbf{Executor}\\Python vs.\ Fortran};
  \node[tag=orange!70!black] at ([xshift=-4pt]generator.north east) {LLM-assisted};
  \node[tag=blue!55!black] at ([xshift=-4pt]translator.north east) {no LLM};
  \node[tag=blue!55!black] at ([xshift=-4pt]executor.north east) {no LLM};

  \draw[flow] (source.east) -- (generator.west);
  \draw[flow] (source.north) -- (0.7,1.55) -- (8.5,1.55) -- (translator.north);
  \draw[flow] (generator.east) -- (library.west);
  \draw[flow] (library.east) -- (translator.west);
  \draw[flow] (translator.east) --
    node[above,font=\scriptsize,align=center] {Python\\module} (executor.west);

  \draw[group] (1.8,-2.35) rectangle (13.75,-4.33);
  \node[legend] at (1.95,-2.35) {Assurance};
  \node[box=3.1cm,llm,minimum height=1.05cm] (auditor) at (5.33,-3.23)
    {\textbf{Auditor}\\library rules};
  \node[artifact=1.7cm] at (8.37,-3.11) {};
  \node[artifact=1.7cm] at (8.31,-3.17) {};
  \node[artifact=1.7cm] (prooflib) at (8.25,-3.23) {\textbf{Proof library}\\Lean proofs};
  \node[box=3.7cm,person,minimum height=1.05cm] (expert) at (11.58,-3.23)
    {\textbf{Human-Expert}\\reviews models and obligations};
  \node[tag=orange!70!black] at ([xshift=-4pt]auditor.north east) {LLM-assisted};
  \node[tag=violet!65!black] at ([xshift=-4pt]expert.north east) {human};
  \draw[flow] (auditor.east) -- (prooflib.west);
  \draw[review] (expert.west) -- (prooflib.east);
  \node[font=\scriptsize\itshape,text=black!70] at (7.775,-4.05)
    {Lean's kernel checks proofs or refutations; Fortran and Python semantics are modeled separately.};

  \draw[feedback,-latex] (executor.south) -- (12.25,-1.7) -- (3.3,-1.7)
    -- (generator.south);
  \draw[feedback] (translator.south) -- (8.5,-1.7);
  \draw[feedback] (auditor.north) -- (5.33,-1.7);
  \fill[red!60!black] (8.5,-1.7) circle (1.4pt) (5.33,-1.7) circle (1.4pt);
  \node[fblabel] at (12.15,-1.375) {testing failures};
  \node[fblabel] at (8.4,-1.375) {unsupported sites};
  \node[fblabel] at (5.23,-2.025) {audit refutes \& counterexamples};
\end{tikzpicture}%
}
  \caption{Design overview of \sys. The Rule-Generator, Translator, and Executor form the minimal core. 
  The Auditor provides an additional feedback channel to provider stronger assurance and capabilities beyond the underlying LLM.}
  \Description{The Rule-Generator, Translator, and Executor share a row
  enclosed as the minimal core, with the Auditor and Human-Expert below. The LLM-assisted
  Rule-Generator creates or revises executable rules in a reusable library.
  The Translator applies these rules deterministically to Fortran frontend facts
  and emits Python modules. The Executor compares native Fortran and translated
  Python using differential and in-situ execution on the target High-Performance Computing (HPC) system.
  The Auditor proposes mechanized semantics, rule obligations, and proof scripts
  with LLM assistance and submits the claims to Lean's kernel. Human-Expert
  review checks semantic-model fidelity and correspondence between obligations
  and executable rules. Unsupported sites, testing failures, audit refutations,
  and theorem premises feed back to the Rule-Generator. }
  \label{fig:approach}
\end{figure}

\sys turns recurring translation decisions into executable rules and audits them against mechanized semantics.
Figure~\ref{fig:approach} presents its design overview, comprising five components connected through a feedback loop.

\sys starts with the \textbf{Rule-Generator} which takes a Fortran program unit to be translated as the input and invokes LLMs to generate executable rules.
It is also assigned with debugging tasks to revise rules in response to testing failures, auditing refutes, and counterexamples from other components in the feedback loop.
Newly generated and revised rules are incorporated into the rule library for reuse in subsequent translations and also enable further iterations of the feedback loop.

Retrieving rules from the library, \textbf{Translator} apply them to Fortran code and produce Python versions. 
Its frontend first parses the Fortran source and extracts program facts, including syntax trees, types, interfaces, constants, and effects. 
The Translator then matches these facts against rules in the library and deterministically emits Python code. 
If matching fails, the Translator will feedback this unsupported site to the Rule-Generator.
The finally generated code is submitted by the \textbf{Executor} to the target High-Performance Computing (HPC) system for execution, and its outputs are compared against those of the native Fortran implementation to identify discrepancies as testing failures. 
Neither the Translator nor the Executor invokes LLMs.

Even if no discrepancy is observed, however, the translation is not necessarily semantically correct, as illustrated by our \texttt{ANINT} example. 
To provide an additional feedback channel for stronger assurance, the \textbf{Auditor} first translates executable rules, together with the relevant Fortran and Python semantic models, into proof obligations in Lean.
The resulting obligations and proof scripts are then submitted to Lean's kernel, which checks proofs or refutations of the modeled claims.
The counterexamples from the LLM-assisted Auditor will be used by the Rule-Generator to repair the executable rules.
As \sys is applied to more scientific software, this feedback loop expands the rule library and progressively develops and refines the semantic models, obligations, and proofs or refutations.

The Auditor invokes LLMs to propose mechanized semantics, rule obligations, and proof scripts, drawing on sources such as compiler and runtime behavior, the Fortran 2018 standard, the Python Language Reference, etc.
Because these LLM-generated artifacts may be incorrect or incomplete, we introduce \textbf{Human-Expert} to examine two critical boundaries: first, whether the semantic models faithfully represent the intended language behavior, and second, whether the generated rule obligations accurately capture the behavior of the corresponding executable rules.
Readers may wonder how Lean obligations can provide assurance about executable rules when both are generated with LLM assistance.
We discuss this question in Section~\ref{sec:obligations}.

Finally, \sys forwards testing failures and audit refutations to the Rule-Generator, which revises executable rules by, for example, strengthening their applicability guards.
The four automated components operate as specialized subagents with fixed workflows, and LLM calls use predefined prompts. 
LLM judgment is confined to local decisions to limit unintended behavior.
The following subsections describe each component in detail.

\subsection{Rule-Generator, Translator, and Executor}
\label{sec:rule-development}

Leaving assurance aside, the Rule-Generator, Translator, and Executor form the minimal core of \sys for turning recurring translation decisions into reusable rules.

In the Rule-Generator, we represent a translation rule as $r=(p_r,g_r,e_r)$, comprising a source pattern $p_r$, an applicability guard $g_r$, and an emission operation $e_r$. 
The source pattern identifies the Fortran construct to be translated. 
The guard constrains the rule using syntax, declared types and kinds, array rank and bounds, procedure interfaces, effects, and numerical configuration. 
The emission operation produces Python expressions, statements, or calls to runtime helpers.

The guard defines the scope over which an emission is claimed to be correct.
In our \texttt{ANINT} example, the counterexample $\texttt{ANINT(2.5)}=3$ versus $\texttt{np.round(2.5)}=2$ refutes the emission itself.
Replacing \texttt{np.round} with a helper that implements rounding ties away from zero repairs this discrepancy, but the rule may be reused only for the argument types, ranks, and \texttt{KIND} variants supported by the helper. 
A repair therefore revises the guard--emission pair, rather than merely replacing a piece of generated code, and the revised rule must be tested and audited again.

Given the same source, analysis facts, configuration, and rule library, the Translator deterministically produces the same Python code. 
It assembles the translated subprograms into a complete module containing constants, companion module imports, persistent module state, signature information, and runtime helpers.
For example, an array reference is shifted using its declared lower bound, while a procedure call uses interface and effect information to propagate changes to \texttt{OUT} and \texttt{INOUT} arguments and shared state.


If the rule library does not cover a Fortran construct, the Translator does not emit a speculative translation. 
Instead, it records the unsupported site, including the enclosing subprogram, source span, original construct, and the reason that no rule applies. 
Each unresolved entry is sent to the Rule-Generator, which invokes an LLM with the program unit, relevant analysis facts, and rejection reason to generate a new guarded rule or revise an existing one. 
Once the resulting rule is added to the rule library, the Translator repeats the deterministic translation pass. The translation remains incomplete until all unresolved entries have been addressed.

The Executor validates the translated Python version against the native Fortran implementation through two complementary approaches. 
The first is differential execution, which runs the native Fortran and translated Python versions on the same inputs. 
This controlled, side-by-side execution provides fine-grained observability and helps attribute a discrepancy to a specific translated routine.
The second is in-situ execution, which retains the original application driver and replaces selected computational routines with their translated versions. 
This hybrid execution reveals integration errors that arise only from procedure calls, argument updates, and shared state.

In both approaches, instrumentation records argument values at procedure entry and exit, including \texttt{IN}, \texttt{OUT}, and \texttt{INOUT} arguments, as well as the values used to evaluate conditional branches. 
The Executor compares the resulting execution traces together with returned values, shared state, and generated files. 
This fine-grained comparison can identify an intermediate divergence even when it is later masked in the final outputs, and can localize the first mismatching procedure or branch condition. 
The resulting diagnostic information is returned to the Rule-Generator to guide rule revision. 
Routines and instrumentation sites that cannot be exercised are explicitly reported as uncovered rather than treated as passing.



\subsection{Auditor and Human-Expert}
\label{sec:obligations}

While the Executor can validate the translated Python version against the native Fortran, this validation is limited to the tested inputs and can miss semantic mismatches.
Even with unlimited feedback iteration in the minimal core, LLM could still retain failures, as our motivating example illustrates.
\sys introduces the Auditor as an additional feedback channel to provide stronger assurance and gets rid of being limited by the capabilities of the underlying LLM.
Here, we address two key questions about how the Auditor works and
what guarantees it provides.

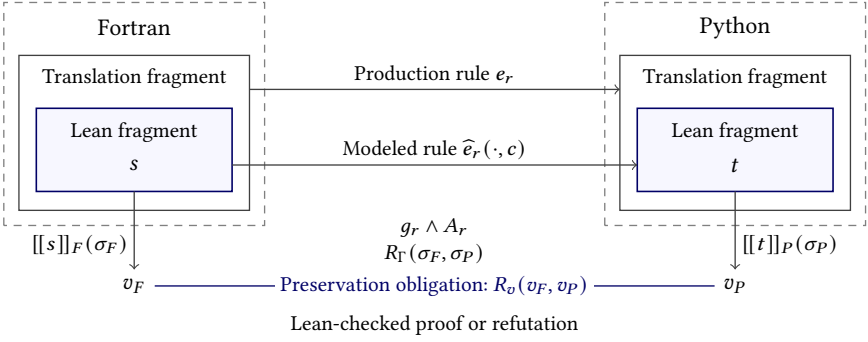
\begin{figure}[t]
  \centering
\cachetikzfigure{assurance-domains}{figs/fig_assurance_domains.tex}
\begin{tikzpicture}[x=1cm,y=1cm,font=\small,
  edge/.style={line width=0.5pt,draw=black!75},
  flow/.style={edge,->},
  model/.style={line width=0.6pt,draw=blue!40!black,fill=blue!3},
  label/.style={font=\footnotesize,inner sep=2pt}]
  \foreach \dx in {0,7.95} {
    \begin{scope}[xshift=\dx cm]
      \draw[edge,dash pattern=on 3pt off 2.5pt,black!50]
        (0,2.35) rectangle (3.45,5.30);
      \draw[edge] (0.20,2.55) rectangle (3.25,4.60);
      \path[model] (0.43,2.80) rectangle (3.02,3.90);
      \node[label] at (1.725,4.30) {Translation fragment};
      \node[label] at (1.725,3.58) {Lean fragment};
    \end{scope}
  }
  \node at (1.725,4.97) {Fortran};
  \node at (9.675,4.97) {Python};
  \node (src) at (1.725,3.15) {$s$};
  \node (tgt) at (9.675,3.15) {$t$};
  \draw[flow] (3.25,4.15) -- node[above,label]
    {Production rule $e_r$} (8.15,4.15);
  \draw[flow] (3.02,3.15) -- node[above,label]
    {Modeled rule $\widehat e_r(\cdot,c)$} (8.38,3.15);

  \node[label,align=center] at (5.70,2.15)
    {$g_r \land A_r$\\$R_\Gamma(\sigma_F,\sigma_P)$};
  \node (fv) at (1.725,1.55) {$v_F$};
  \node (pv) at (9.675,1.55) {$v_P$};
  \draw[flow] (1.725,2.80) -- node[pos=0.68,left,label]
    {$[\![s]\!]_F(\sigma_F)$} (fv.north);
  \draw[flow] (9.675,2.80) -- node[pos=0.68,right,label]
    {$[\![t]\!]_P(\sigma_P)$} (pv.north);
  \draw[edge,blue!40!black] (fv.east) -- (pv.west);
  \node[label,fill=white,text=blue!40!black] at (5.70,1.55)
    {Preservation obligation: $R_v(v_F,v_P)$};
  \node[label] at (5.70,1.03) {Lean-checked proof or refutation};
\end{tikzpicture}
  
  \caption{Nested scopes of translation and semantic assurance.
  Full Fortran and Python lack complete mechanized semantic ground truth.
  Translation rules address a fragment of each language.
  Within each translation fragment, our Lean development models a smaller subfragment.}
  \Description{Two matching nested regions show full Fortran and Python
  languages, their translation fragments, and smaller Lean fragments.
  Production and modeled rules connect the two sides horizontally. Independent
  semantic evaluations descend from the source and target expressions to their
  results. The bottom relation is the preservation obligation, subject to the
  guard, premises, and related initial states shown in the center.}
  \label{fig:assurance-domains}
\end{figure}

\vspace{5pt}
\noindent \textbf{What does the Auditor assure?}
Figure~\ref{fig:assurance-domains} illustrates three nested scopes on each side of Fortran and Python: the full language, the fragment addressed by executable translation rules, and the innermost fragment modeled in Lean.
To the best of our knowledge, no complete formal semantics is available for either full Fortran or full Python.
To construct the innermost Lean fragment, \sys mechanizes selected constructs based on language specifications, including the Fortran 2018 standard, the Python Language Reference, and NumPy documentation, supplemented by compiler and runtime behavior and concrete probes.

For a rule $r=(p_r,g_r,e_r)$, let $s$ match the source pattern $p_r$ in static context $c$, including $\Gamma$, and let $\widehat e_r(s,c)$ denote the modeled translation. The expression obligation universally quantifies the following implication over matching expressions, contexts, states, and results:
\begin{equation}
\label{eq:rule-obligation}
\begin{aligned}
 &g_r(s,c) \land A_r(s,c,\sigma_F,\sigma_P)
       \land R_\Gamma(\sigma_F,\sigma_P)
       \land [\![s]\!]_F(\sigma_F)=v_F\\
 &\quad\Longrightarrow
       \exists v_P.\;[\![\widehat e_r(s,c)]\!]_P(\sigma_P)=v_P
       \land R_v(v_F,v_P).
\end{aligned}
\end{equation}
Here, $A_r$ includes modeled well-formedness conditions and rule-specific premises, and $R_v$ relates source and target results.
The obligation states that, whenever the guard and premises hold, the initial states correspond, and source evaluation yields a value, target evaluation must also yield a related value.
The obligation applies only when the modeled Fortran expression has a defined result (\eg, not division by zero).
The resulting guarantee is therefore a forward preservation property over source-defined executions within the modeled fragments, not a bidirectional equivalence between the full languages.
Our current Lean semantics model arithmetic instead of exactly the IEEE 754 floating-point behavior.

The Auditor's checked proofs establish preservation within the innermost Lean fragment under explicit guards and premises.
Counterexamples, as another form of assurance, supplied to the Rule-Generator reveal conditions under which executable rules fail.
The Rule-Generator then corrects the emission or strengthens the applicability guard.
These counterexamples guide model extension and rule refinement, progressively bringing additional rules within the modeled scope and supporting renewed attempts by the LLM to construct preservation proofs for Lean to check.


\vspace{5pt}
\noindent \textbf{How can the LLM-generated Auditor provide assurance for LLM-generated rules?}
In the Auditor, the mechanized semantic definitions and Lean proof obligations are not handcrafted but automatically generated by invoking LLMs, just like the executable rules.
However, the Auditor can still provide assurance for executable rules. 

First, an executable rule and its proof obligation serve different roles.
The rule proposes a transformation, while the obligation asks whether it preserves behavior under independently defined Fortran and Python semantics.
An obligation can therefore faithfully represent an incorrect rule and expose a false preservation claim.

Second, obligations are universally quantified and checked through symbolic reasoning, extending assurance beyond finite execution-based tests.
A valid proof establishes preservation for every modeled instance satisfying the guard and premises, while a single checked counterexample refutes the universal claim.
For example, the obligation for translating \texttt{ANINT} to \texttt{np.round} is refuted at \texttt{x = 2.5}, where the source model returns $3$ and the target model returns $2$.

Third, human review of two correspondences are required before applying audit results to production rules.
Experts compare the mechanized source and target semantics against language specifications, compiler and runtime behavior, and concrete probes to make sure that the semantic models capture the intended language behavior.
They also verify that each obligation's modeled guard, emission, and runtime helpers are consistent with the executable rule's implementation.

\subsection{Implementation}
\label{sec:implementation}

The \sys prototype builds on \texttt{fparser} and Lean~4 with Mathlib version~4.34.0.
It contains approximately 50,000 lines of Python code for implementing each component and integration.

The prototype invokes LLM coding agents through their non-interactive command-line interfaces. 
Specifically, it launches Claude Code using \texttt{claude -p} and invokes Codex through its analogous headless command, with both tools authenticated through user subscriptions.
For each invocation, \sys constructs a task-specific prompt and starts the agent in the relevant workspace, allowing it to inspect existing artifacts, modify designated files, and execute supporting tools. 
The resulting file changes and process outputs are then collected by \sys for compilation, execution, or Lean validation.
Any diagnostics can be incorporated into a subsequent invocation for refinement.

In the Rule-Generator, the library of the rule $r=(p_r,g_r,e_r)$ is
implemented as expression and statement handlers, intrinsic mappings, array-index transformations, procedure-call transformations, and module construction, as shown in the motivating example and more in following sections.
Runtime helpers implement target-language operations that cannot be emitted directly as Python expressions.

The Executor observes program state through generated calling wrappers and lightweight runtime probes. 
For differential execution, it uses interface information extracted by \texttt{fparser} to generate flat \texttt{f2py} wrappers around the original Fortran procedures. 
It constructs one input instance, passes independent copies to the compiled Fortran procedure and translated Python procedure, and collects their returned values and modified argument buffers. 
This captures \texttt{INTENT(IN)} arguments before execution, \texttt{INTENT(OUT)} arguments after execution, and both states of \texttt{INTENT(INOUT)} arguments. 
For in-situ execution, explicitly inserted runtime probes emit named checksums and numerical summaries of selected arrays at corresponding execution points.
The Executor aligns these observations by procedure or probe identifier and compares them using bit-exact and ULP-based tolerance criteria.

The Auditor traces each obligation with its corresponding production rule and implementation locations. 
Each entry records the proof status, stated assumptions, applicability conditions, and generated examples used during review.
Validation scripts compile the Lean development, detect unfinished proofs, and inspect the axioms on which recorded theorems depend.

\section{LLM-Generated Components in \sys}
\label{sec:analysis}

Being applied to 15 real-world scientific software (Section~\ref{sec:evaluation}), \sys has accumulated executable rules, semantic definitions, and proofs that are reusable and also form a critical part of the \sys system.
This section describes their organization, as well as the connections between the executable and proof components.

\begin{figure}[t]
\centering
\begingroup

\newsavebox{\duoExampleBox}
\begin{lrbox}{\duoExampleBox}
\begin{minipage}[t]{0.60\linewidth}
\centering
\normalsize\textbf{Translation Example}
\vspace{5pt}
\begin{lstlisting}[style=casecode,
  basicstyle=\normalsize\ttfamily,
  morekeywords={exp,size,do,enddo,for,range,math,_f_powi},
  frame=tb, framesep=3pt,
  rulecolor=\color{black}, framerule=\heavyrulewidth,
  xleftmargin=0pt, xrightmargin=0pt,
  framextopmargin=0pt, framexbottommargin=0pt,
  aboveskip=6pt, belowskip=0pt, lineskip=-1.5pt]
// Fortran source
// Real scalars, integer k and N
// One-based array parameters
N = size(parameters)
r0 = parameters(1)
beta = parameters(2)
z = (r-r0)*exp(-beta*(r-r0)**2)
f = 0
do k = 0, N-3
  f = f + parameters(k+3)*z**k
enddo

// Python translation
n = np.size(parameters)
r0 = parameters[0]
beta = parameters[1]
z = ((r-r0) *
     math.exp(-beta * ((r-r0) * (r-r0))))
f = 0
for k in range(0, n-3+1):
    f = (f +
         parameters[(k+3)-1] * _f_powi(z, k))
\end{lstlisting}
\vspace{-3pt}
\end{minipage}
\end{lrbox}

\newsavebox{\duoRulesBox}
\begin{lrbox}{\duoRulesBox}
\begin{minipage}[t]{0.60\linewidth}
\centering
\normalsize\textbf{Executable Rules}
\vspace{5pt}
\begin{lstlisting}[style=casecode,
  basicstyle=\normalsize\ttfamily,
  morekeywords={def,if,return,expand_power,integer_literal,is_integer},
  frame=tb, framesep=3pt,
  rulecolor=\color{black}, framerule=\heavyrulewidth,
  xleftmargin=0pt, xrightmargin=0pt,
  framextopmargin=0pt, framexbottommargin=0pt,
  aboveskip=6pt, belowskip=0pt, lineskip=-1.5pt]
// Guarded power rules
def emit_power(x, n, n_ast):
    p = integer_literal(n_ast)
    if p is not None and p != 0:
        return expand_power(x, p)
    if is_integer(n_ast):
        return f"_f_powi({x}, {n})"

// Dispatch tables
ARITH_OPS = {"+": "+", "-": "-", "*": "*"}
scalar_table = {"exp": "math.exp",
                "size": "np.size"}
def emit_intrinsic(name, x):
    return f"{scalar_table[name]}({x})"

// One-based array indexing
def emit_index(a, i):
    return f"{a}[({i}) - 1]"

// Unit-step counted loop
def emit_do(k, lo, hi):
    return f"for {k} in range({lo}, {hi}+1):"
\end{lstlisting}
\vspace{-3pt}
\end{minipage}
\end{lrbox}

\newsavebox{\duoCombinedBox}
\begin{lrbox}{\duoCombinedBox}
\begin{minipage}[t]{\wd\duoExampleBox}
\usebox{\duoExampleBox}
\end{minipage}
\hspace{10pt}
\begin{minipage}[t]{\wd\duoRulesBox}
\usebox{\duoRulesBox}
\end{minipage}
\end{lrbox}
\resizebox{\linewidth}{!}{\usebox{\duoCombinedBox}}
\endgroup

\caption{Rule composition in Duo's \texttt{dipole\_polynom\_exp}.
Left: a Fortran fragment and its generated Python translation, with hoisted
constants and redundant parentheses simplified.
Right: selected rules shown as simplified emitter pseudocode for scalar
values, one-based arrays, unit-step loops, and the expanded-power profile.
The constant square expands into multiplication, while the variable integer
power calls \texttt{\_f\_powi}. Intrinsic dispatch, index rebasing, and inclusive
loop bounds compose with arithmetic and assignment rules.}
\Description{Two code panels show a polynomial with exponential decay in
Fortran and Python, alongside its reusable translation rules. A constant
square becomes multiplication, a variable integer power calls a shared
runtime helper, array indices shift by one, and the counted loop retains
its inclusive upper bound.}
\label{fig:rule-composition}
\end{figure}

\subsection{Executable Rule Library}
\label{sec:analysis-rules}

The executable library contains 193 
executable rules, which comprises 176 language or numerical transformations and 17 environment or extension-interface rules.
Some rules cover a family of operations. 
Table~\ref{tab:registered-rules} in Appendix groups them by their categories.
The largest category concerns precision and kind policy, followed by control flow and environment interfaces.
Other categories cover procedure calls, arrays, indexing, memory operations, arithmetic, reductions, and I/O.


\vspace{5pt}
\noindent \textbf{Guarded branches, dispatch tables, and helpers.} 
Figure~\ref{fig:rule-composition} illustrates these mechanisms using an example in translating \texttt{dipole\_polynom\_exp} subroutine in Duo~\cite{yurchenko2016duo}, a software calculating diatomic molecules by solving the nuclear motion Schrödinger equation.
The guarded power emitter checks the exponent and operand types before selecting an emission.
For \texttt{(r-r0)**2}, \texttt{integer\_literal(n\_ast)} returns the nonzero constant 2, so the first branch calls \texttt{expand\_power} to produce \texttt{(r-r0)*(r-r0)}. 
This expansion deliberately matches the reference Fortran compiler's lowering of the constant square to multiplication to preserve bitwise-identical, since CPython's floating-point \texttt{**} may use the platform \texttt{pow}, whose rounding can differ from multiplication~\cite{gcc2026fortranpower,cpython2026floatpower,gcc2024builtins}. 
For \texttt{z**k}, the exponent \texttt{k} has integer type but no value known at translation time. 
The second branch therefore emits \texttt{\_f\_powi(z,k)}. 
The \texttt{exp} call in the assignment to \texttt{z} uses a dispatch table. 
\texttt{emit\_intrinsic} looks up "exp" in \texttt{scalar\_table}, obtains \texttt{math.exp}, and emits a call with the translated argument. 
The generated \texttt{\_f\_powi(z,k)} call invokes a shared runtime helper that evaluates integer powers by repeated squaring in gfortran’s multiplication order. 
This behavior is implemented once in the helper and reused by every generated call site.

\begin{figure}[t]
\centering
\begingroup
\newlength{\duoMappingWidth}
\newlength{\duoTreeWidth}
\setlength{\duoMappingWidth}{0.55\linewidth}
\setlength{\duoTreeWidth}{0.42\linewidth}

\newcommand{\duoMappingRule}[2]{\raggedright\textbf{#1.}\ {\fontsize{8}{9.6}\selectfont\texttt{#2}}}
\newsavebox{\duoRuleMappingTableBox}
\begin{lrbox}{\duoRuleMappingTableBox}
\begin{minipage}{\duoMappingWidth}
\fontsize{8.5}{10.2}\selectfont
\setlength{\tabcolsep}{3pt}
\renewcommand{\arraystretch}{1.12}
\begin{tabular}{@{}p{0.44\linewidth}p{\dimexpr0.56\linewidth-6pt\relax}@{}}
\toprule
\textbf{Rule} & \textbf{Fortran $\to$ Python} \\
\midrule
\duoMappingRule{R1}{R-power.expanded} &
\texttt{(r-r0)**2}\newline
$\to$ \texttt{(r-r0)*(r-r0)} \\[3pt]
\duoMappingRule{R2}{R-intrinsic.\allowbreak scalar\_\allowbreak table} &
\texttt{exp(...)}\newline
$\to$ \texttt{math.exp(...)} \\[3pt]
\duoMappingRule{R3}{R-power.\allowbreak integer\_\allowbreak runtime} &
\texttt{z**k}\newline
$\to$ \texttt{\_f\_powi(z,k)} \\[3pt]
\duoMappingRule{R4}{R-index.shift\_\allowbreak unit} &
Subscript \texttt{k+3}\newline
$\to$ \texttt{(k+3)-1} \\[3pt]
\duoMappingRule{R5}{R-index.array\_\allowbreak reference} &
\texttt{parameters(k+3)}\newline
$\to$ \texttt{parameters[(k+3)-1]} \\[3pt]
\duoMappingRule{R6}{R-arithmetic.\allowbreak scalar\_\allowbreak fallback} &
\texttt{r-r0} $\to$ \texttt{(r-r0)}\newline
\texttt{k+3} $\to$ \texttt{(k+3)}\newline
\texttt{f+...} $\to$ \texttt{(f+...)} \\[3pt]
\duoMappingRule{R7}{R-assignment.\allowbreak ordinary} &
\texttt{f = ...}\newline
$\to$ \texttt{f = ...} \\[3pt]
\duoMappingRule{R8}{R-loop.counted} &
\texttt{do k=0,N-3}\newline
$\to$ \texttt{for k in}\newline
\hspace*{1em}\texttt{range(0,n-3+1):} \\
\bottomrule
\end{tabular}
\end{minipage}
\end{lrbox}
\newlength{\duoRuleMappingTableHeight}
\setlength{\duoRuleMappingTableHeight}{\dimexpr
  \ht\duoRuleMappingTableBox+\dp\duoRuleMappingTableBox\relax}

\def\duoTreeYScale{1}
\newcommand{\duoCompositionTree}{%
\begin{tikzpicture}[
  yscale=\duoTreeYScale,
  rule/.style={draw=black!45, rounded corners=2pt, fill=white,
    align=center, font=\fontsize{7.5}{8.5}\selectfont\sffamily,
    inner xsep=4pt, inner ysep=3pt},
  leaf/.style={draw=black!45, rounded corners=2pt, fill=white,
    align=center, font=\fontsize{7.5}{8.5}\selectfont\ttfamily,
    inner xsep=6pt, inner ysep=3pt},
  edge/.style={->, draw=black!65, line width=0.5pt},
  edge label/.style={font=\fontsize{6}{7}\selectfont\sffamily,
    fill=white, inner sep=1pt}
]
\node[rule, text width=2.4cm] (loop) at (0.4,0)
  {\textbf{R8} Counted loop\\
   {\fontsize{7}{8}\selectfont\ttfamily range(0,n-3+1)}};
\node[rule, text width=1.5cm] (bound) at (-1,-1)
  {\textbf{R6} Subtract\\\texttt{n-3}};
\node[rule, text width=1.5cm] (assign) at (1,-1)
  {\textbf{R7} Assign\\\texttt{f = ...}};
\node[rule, text width=1.5cm] (add) at (1,-2)
  {\textbf{R6} Add\\\texttt{f + ...}};
\node[leaf] (f) at (-0.8,-3) {f};
\node[rule, text width=1.7cm] (multiply) at (1.8,-3)
  {\textbf{R6} Multiply\\\texttt{... * ...}};
\node[rule, text width=1.7cm] (array) at (0.25,-4)
  {\textbf{R5} Array ref.\\
   {\fontsize{6.5}{7.5}\selectfont\ttfamily parameters[...]}};
\node[rule, text width=1.7cm] (power) at (2.5,-4)
  {\textbf{R3} Int. power\\
   {\fontsize{7}{8}\selectfont\ttfamily \_f\_powi(z,k)}};
\node[rule, text width=1.7cm] (shift) at (0.25,-5)
  {\textbf{R4} Rebase\\\texttt{(k+3)-1}};
\node[rule, text width=1.5cm] (index) at (0.25,-6)
  {\textbf{R6} Add\\\texttt{k+3}};
\draw[edge] (loop.south) -- node[edge label, midway, left]
  {bound} (bound.north);
\draw[edge] (loop.south) -- node[edge label, midway, right]
  {body} (assign.north);
\draw[edge] (assign.south) -- node[edge label, midway, right]
  {rhs} (add.north);
\draw[edge] (add.south) -- (f.north);
\draw[edge] (add.south) -- (multiply.north);
\draw[edge] (multiply.south) -- (array.north);
\draw[edge] (multiply.south) -- (power.north);
\draw[edge] (array.south) -- node[edge label, midway, right]
  {subscript} (shift.north);
\draw[edge] (shift.south) -- (index.north);
\end{tikzpicture}%
}
\newsavebox{\duoCompositionTreeMeasureBox}
\sbox{\duoCompositionTreeMeasureBox}{\duoCompositionTree}
\pgfmathsetmacro{\duoTreeYScale}{1 +
  ((\the\wd\duoCompositionTreeMeasureBox / \the\duoTreeWidth) *
    \the\duoRuleMappingTableHeight -
    (\the\ht\duoCompositionTreeMeasureBox +
     \the\dp\duoCompositionTreeMeasureBox)) / (6cm)}
\newsavebox{\duoCompositionTreeBox}
\sbox{\duoCompositionTreeBox}{\resizebox{\duoTreeWidth}{!}{\duoCompositionTree}}

\newsavebox{\duoMappingCaptionBox}
\begin{lrbox}{\duoMappingCaptionBox}
\begin{minipage}[t]{\duoMappingWidth}
\captionsetup*{type=table,position=top}
\caption{Executable rules and representative Fortran-to-Python fragments in
Figure~\ref{fig:rule-composition}. Labels identify nodes in Figure~\ref{fig:rule-composition-tree}.}
\label{tab:rule-composition-mapping}
\end{minipage}
\end{lrbox}
\newsavebox{\duoTreeCaptionBox}
\begin{lrbox}{\duoTreeCaptionBox}
\begin{minipage}[t]{\duoTreeWidth}
\captionsetup*{type=figure,position=top}
\caption{Composition tree for the counted loop in Figure~\ref{fig:rule-composition}.
Repeated labels denote separate rule applications.}
\label{fig:rule-composition-tree}
\end{minipage}
\end{lrbox}
\newlength{\duoCaptionRowHeight}
\setlength{\duoCaptionRowHeight}{\dimexpr
  \ht\duoMappingCaptionBox+\dp\duoMappingCaptionBox\relax}
\ifdim\duoCaptionRowHeight<\dimexpr\ht\duoTreeCaptionBox+\dp\duoTreeCaptionBox\relax
  \setlength{\duoCaptionRowHeight}{\dimexpr
    \ht\duoTreeCaptionBox+\dp\duoTreeCaptionBox\relax}
\fi
\newsavebox{\duoMappingPanelBox}
\sbox{\duoMappingPanelBox}{\vbox{\offinterlineskip
  \vbox to\duoCaptionRowHeight{\hbox{\usebox{\duoMappingCaptionBox}}\vfil}
  \vskip6pt
  \hbox{\usebox{\duoRuleMappingTableBox}}}}
\newsavebox{\duoTreePanelBox}
\sbox{\duoTreePanelBox}{\vbox{\offinterlineskip
  \vbox to\duoCaptionRowHeight{\hbox{\usebox{\duoTreeCaptionBox}}\vfil}
  \vskip6pt
  \hbox{\usebox{\duoCompositionTreeBox}}}}
\begin{minipage}[t]{\duoMappingWidth}
\vspace{0pt}\noindent\usebox{\duoMappingPanelBox}
\end{minipage}\hfill
\begin{minipage}[t]{\duoTreeWidth}
\vspace{0pt}\noindent\usebox{\duoTreePanelBox}
\end{minipage}
\endgroup
\Description{A two-column table on the left lists eight executable rules, with
bold labels R1 through R8 merged into the rule names and arrows connecting
representative Fortran and Python fragments. An equally tall tree on the right
shows the counted loop, upper bound, assignment, addition, multiplication,
array reference, integer power, index rebasing, and subscript addition.}
\end{figure}

\vspace{5pt}
\noindent \textbf{Chaining rules in a tree structure.}
We list rules applied in translating this example in Table~\ref{tab:rule-composition-mapping}. 
Figure~\ref{fig:rule-composition-tree} shows how these rules are chained in a tree structure for translation.
The root applies R8 to the loop, with separate branches for the upper-bound expression \texttt{N-3}, handled by R6, and the body assignment, handled by R7.
Within the assignment, R6 translates the index expression \texttt{k+3}, R4 rebases it to \texttt{(k+3)-1}, and R5 emits the array reference \texttt{parameters[(k+3)-1]}.
A separate branch applies R3 to \texttt{z**k}, emitting \texttt{\_f\_powi(z,k)} because \texttt{k} is a runtime integer exponent.
The enclosing applications of R6 combine the array element and the translated power into a product, then add that product to \texttt{f}.
R7 emits the assignment, and R8 combines the translated bounds and body into the Python loop, adding one to the upper bound to preserve Fortran's inclusive iteration range.
Each occurrence of R6 represents a separate application of the same arithmetic rule.
The composition follows the nesting of source constructs, enclosing rules assemble their output from translated children, allowing independently rules to translate the complete loop.

\begingroup
\definecolor{proofHBlue}{HTML}{416D95}
\definecolor{proofHGreen}{HTML}{397D70}
\definecolor{proofHOrange}{HTML}{C77B23}
\definecolor{proofHBorder}{HTML}{A8B7C5}
\definecolor{proofHFill}{HTML}{F0F4F8}
\definecolor{proofHFamily}{HTML}{EDF6F2}

\normalsize
\newlength{\proofExampleRefWidth}
\settowidth{\proofExampleRefWidth}{\textbf{\texttt{np.copysign}}}
\setlength{\proofExampleRefWidth}{\dimexpr5\proofExampleRefWidth\relax}
\lstset{style=styleD,language=LeanD,numbers=none,
  xleftmargin=0pt,xrightmargin=0pt,
  rulecolor=\color{black},framerule=\heavyrulewidth}
\newsavebox{\proofLayersOneTwoBox}
\begin{lrbox}{\proofLayersOneTwoBox}
\begin{minipage}[t]{\proofExampleRefWidth}
\raggedright\small
\hrule height\heavyrulewidth
\vspace{8pt}
\textbf{\textcolor{proofHBlue}{I}: Syntax and semantics}
\begin{lstlisting}[frame=b]
-- F.eval: binary-expression case
| .bin op a b => do
    let va ← eval o (*@$\sigma$@*) a
    let vb ← eval o (*@$\sigma$@*) b
    lift2 (arith op) va vb

-- F.Exec: assignment constructor
| assign {(*@$\sigma$@*) (*@$\sigma$@*)' x subs e v} :
    eval o (*@$\sigma$@*) e = some v →
    assign o (*@$\Gamma$@*) (*@$\sigma$@*) x subs v = some (*@$\sigma$@*)' →
    Exec o (*@$\Gamma$@*) (.assign x subs e)
      (*@$\sigma$@*) (*@$\sigma$@*)' .normal
\end{lstlisting}
\par\vspace{3pt}
\textbf{\textcolor{proofHBlue}{II}: Cross-language relations}
\begin{lstlisting}[frame=b]
-- RelS real case: Duo's f=0
| real {kind v w} : P.toR w = some v →
    RelS (.real kind v) w

-- Duo: parameters(k+3) / parameters[(k+3)-1]
structure RelA (a : FArr) (p : PArr) : Prop where
  shape : p.shape = a.extents
  elems : ∀ idx, a.inBounds idx →
    RelS (a.get idx) (p.get (rebase a.lbs idx))

def RelState ((*@$\Gamma$@*) : F.Env)
    ((*@$\sigma$@*) : F.State) ((*@$\tau$@*) : P.State) : Prop :=
  ∀ x d, (*@$\Gamma$@*) x = some d →
    ∃ v w, (*@$\sigma$@*) x = some v ∧ (*@$\tau$@*) x = some w ∧
      RelV v w
\end{lstlisting}
\end{minipage}
\end{lrbox}
\def\proofHYScale{1}
\newcommand{\proofHierarchyDiagram}{%
  \pgfmathsetlengthmacro{\proofHFamilyHeight}{2.35cm*\proofHYScale}
\begin{tikzpicture}[
  yscale=\proofHYScale,
  font=\fontsize{8.5}{10}\selectfont,
  layer/.style={draw=proofHBorder,fill=proofHFill,rounded corners=1.5pt,
    text width=6.05cm,minimum height=0.82cm,align=center,inner sep=3pt,line width=0.4pt},
  example/.style={draw=proofHBorder,fill=white,rounded corners=1pt,
    align=center,inner sep=2pt,line width=0.35pt},
  derivation/.style={->,>=stealth,draw=proofHBlue,line width=0.7pt},
  reuse/.style={->,>=stealth,draw=proofHGreen,dashed,line width=0.7pt},
  composition/.style={->,>=stealth,draw=proofHOrange,line width=0.7pt}
]
\node[layer] (semantics) at (0,0) {
  \textbf{Syntax and semantics}\\[-1pt]
  \texttt{F.eval; P.eval; F.Exec; P.Exec}};
\node[layer] (relations) at (0,1.12) {
  \textbf{Cross-language relations}\\[-1pt]
  \texttt{RelS; RelA; RelState}};
\node[layer,minimum height=1.0cm] (contracts) at (0,2.30) {
  \textbf{Modeled translation and contracts}\\[-1pt]
  \texttt{Tr.trans; SoundScalar; SoundPair}\\[-1pt]
  \texttt{StmtSound; Refuted}};

\node[layer,fill=proofHFamily,minimum height=\proofHFamilyHeight] (families) at (0,4.25) {};
\node[font=\fontsize{8.5}{10}\selectfont\bfseries] at (0,5.18) {Rule-family theorems};
\draw[proofHBorder,line width=0.35pt] (0,3.30) -- (0,4.88);
\node[font=\fontsize{8}{9}\selectfont\bfseries] at (-1.56,4.83) {Rule derivation};
\node[font=\fontsize{8}{9}\selectfont\bfseries] at (1.56,4.83) {Shared-lemma siblings};
\node[example,text width=2.65cm,minimum height=0.47cm] (derived) at (-1.56,4.22) {Derived theorem};
\node[example,text width=2.65cm,minimum height=0.47cm] (basic) at (-1.56,3.40) {Basic theorem};
\draw[derivation] (basic.north) -- (derived.south);
\node[example,text width=1.12cm,font=\fontsize{8}{9}\selectfont,minimum height=0.59cm] (division) at (0.83,4.22) {\texttt{R0017}\\Division};
\node[example,text width=1.12cm,font=\fontsize{8}{9}\selectfont,minimum height=0.59cm] (modulo) at (2.29,4.22) {\texttt{R0077}\\\texttt{MOD}};
\node[example,text width=2.65cm,font=\fontsize{8}{9}\selectfont,minimum height=0.47cm] (lemma) at (1.56,3.40) {\texttt{truncR\_div\_intCast}};
\draw[reuse] (lemma.north) -- (division.south);
\draw[reuse] (lemma.north) -- (modulo.south);

\node[layer,minimum height=1.0cm] (compose) at (0,6.05) {
  \textbf{Composition theorems}\\[-1pt]
  \texttt{trans\_sound; transStmt\_sound}\\[-1pt]
  {\fontsize{8}{9}\selectfont 34 expression rules; 7 statement families}};
\node[layer,minimum height=0.87cm] (kernel) at (0,7.25) {
  \textbf{Whole-kernel theorems}\\[-1pt]
  \texttt{saxpy\_trans; saxpy\_sound}};

\node[anchor=east,text=proofHBlue,font=\bfseries] at (-3.30,0) {I};
\node[anchor=east,text=proofHBlue,font=\bfseries] at (-3.30,1.12) {II};
\node[anchor=east,text=proofHBlue,font=\bfseries] at (-3.30,2.30) {III};
\node[anchor=east,text=proofHBlue,font=\bfseries] at (-3.30,4.25) {IV};
\node[anchor=east,text=proofHBlue,font=\bfseries] at (-3.30,6.05) {V};
\node[anchor=east,text=proofHBlue,font=\bfseries] at (-3.30,7.25) {VI};
\draw[derivation] (semantics.north) -- (relations.south);
\draw[derivation] (relations.north) -- (contracts.south);
\draw[derivation] (contracts.north) -- (families.south);
\draw[composition] (families.north) -- (compose.south);
\draw[composition] (compose.north) -- (kernel.south);

\begin{scope}[yscale={1/\proofHYScale}]
\draw[derivation] (-3.27,-0.86) -- (-2.82,-0.86);
\node[anchor=west,font=\fontsize{7.5}{9}\selectfont] at (-2.77,-0.86) {Derivation};
\draw[reuse] (-1.01,-0.86) -- (-0.56,-0.86);
\node[anchor=west,font=\fontsize{7.5}{9}\selectfont] at (-0.51,-0.86) {Shared lemma};
\draw[composition] (1.46,-0.86) -- (1.91,-0.86);
\node[anchor=west,font=\fontsize{7.5}{9}\selectfont] at (1.96,-0.86) {Composition};
\end{scope}
\end{tikzpicture}%
}
\newsavebox{\proofHierarchyMeasureBox}
\sbox{\proofHierarchyMeasureBox}{\proofHierarchyDiagram}
\pgfmathsetmacro{\proofHYScale}{1 +
  ((\the\wd\proofHierarchyMeasureBox / \the\proofExampleRefWidth) *
    (\the\ht\proofLayersOneTwoBox + \the\dp\proofLayersOneTwoBox) -
    (\the\ht\proofHierarchyMeasureBox + \the\dp\proofHierarchyMeasureBox)) /
  (7.25cm)}

\begin{figure}[!tp]
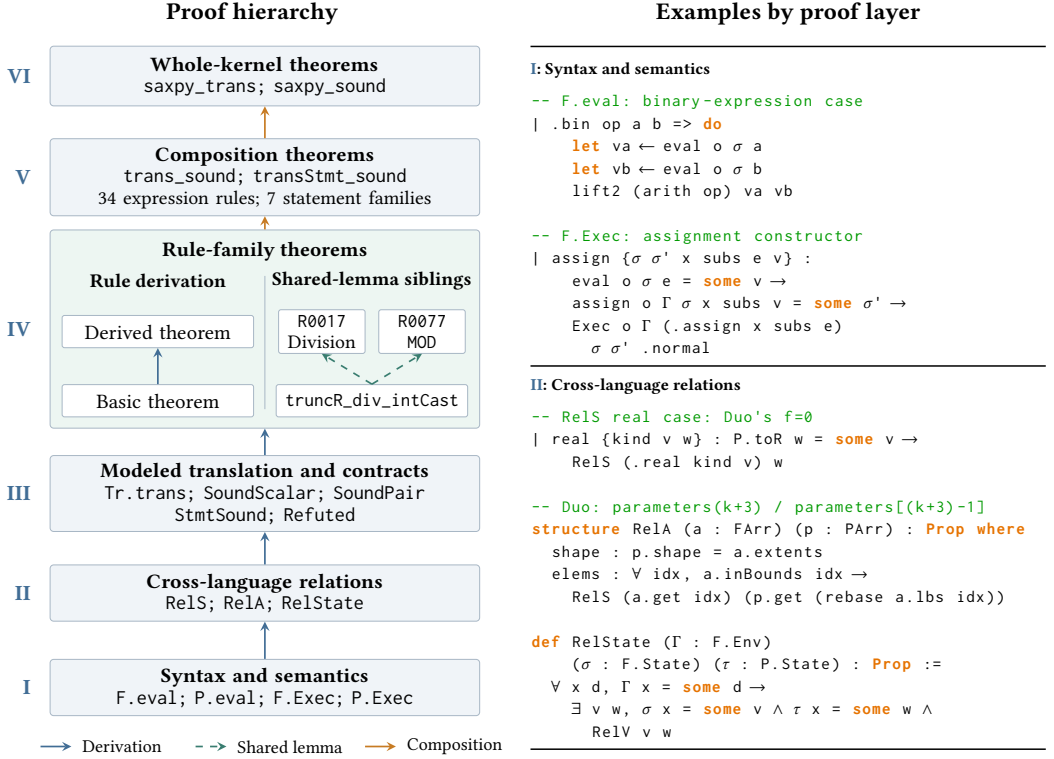

\centering
\begin{minipage}[t]{0.49\linewidth}
\vspace{0pt}\centering
\small\textbf{Proof hierarchy}\par\vspace{8pt}
\nointerlineskip
\resizebox{\linewidth}{!}{%
\proofHierarchyDiagram%
}
\end{minipage}\hfill
\begin{minipage}[t]{0.49\linewidth}
\vspace{0pt}\centering
\small\textbf{Examples by proof layer}\par\vspace{8pt}
\nointerlineskip
\resizebox{\linewidth}{!}{\usebox{\proofLayersOneTwoBox}}
\end{minipage}

\caption{The Lean proof-library hierarchy (left) and representative definitions for Layers I--II (right). Layer I shows tailored snippets from \texttt{F.eval} and \texttt{F.Exec}. Layer II shows scalar, array, and state relations illustrated with Duo example.}
\Description{The left panel shows six proof layers from syntax and semantics through cross-language relations, translation contracts, local rule theorems, composition theorems, and whole-kernel theorems. The right panel arranges Layers I and II from top to bottom, showing Fortran expression evaluation and statement execution, followed by scalar, array, and state relations for Duo.}
\label{fig:proof-hierarchy}
\end{figure}
\endgroup

\subsection{Semantic Models and Proof Library}
\label{sec:analysis-proofs}

The Lean proof library comprises 42 modules and 190 distinct theorem declarations. 
Table~\ref{tab:lean-proof-categories} in Appendix~\ref{app:lean-proof-categories} groups the proofs by semantic family and proof role, with representative theorem declarations.
Executable rules, audit records, and theorem declarations are not in one-to-one correspondence.
A rule's guarded branches may require separate audit records, while one record may cover an emitter together with its runtime helpers.
Each record may cite several theorems, and the same theorem or generic lemma may support multiple records.

Therefore, the library is organized hierarchically into six layers, with definitions and theorems in lower layers supporting proofs in higher layers.
Figure~\ref{fig:proof-hierarchy} shows the hierarchy and representative definitions for Layers I--II, while Figure~\ref{fig:proof-layer-examples} presents contracts and theorems for Layers III--VI.

\vspace{5pt}
\noindent \textbf{Layer I: Syntax and semantics.}
The \texttt{Syntax} modules define separate Abstract Syntax Trees (ASTs) for the modeled Fortran and Python fragments.
Figure~\ref{fig:proof-hierarchy} shows the binary-expression case of \texttt{F.eval} and the assignment constructor of \texttt{F.Exec}.
\texttt{F.eval} is a structurally recursive Lean definition.
In the \texttt{.bin op a b} case, its two recursive calls evaluate \texttt{a} and \texttt{b} in the current state $\sigma$, producing \texttt{va} and \texttt{vb}.
The call \texttt{lift2 (arith op) va vb} then applies the scalar arithmetic operation, with elemental lifting for array operands.
In Duo's \texttt{z**k} example, \texttt{op} is \texttt{.pow}, and \texttt{a} and \texttt{b} are the variable expressions for \texttt{z} and \texttt{k}.
The \texttt{Option} return type represents a defined result as \texttt{some v} and propagates \texttt{none} when an operand evaluation or operation is undefined.

\texttt{F.Exec} is an inductively defined big-step relation connecting a statement, its initial and final states, and a control outcome.
The displayed assignment constructor has two premises.
The first requires the right-hand side \texttt{e} to evaluate to \texttt{some v} in $\sigma$.
The second requires the assignment helper, using declarations $\Gamma$, to update the destination specified by \texttt{x} and \texttt{subs} and produce \texttt{some} $\sigma'$.
The conclusion records that \texttt{.assign x subs e} takes $\sigma$ to $\sigma'$ with the outcome \texttt{.normal}.
The corresponding \texttt{P.eval} and \texttt{P.Exec} independently define target-language evaluation and execution.
Both models use mathematical integers and reals, while arrays retain their dimensions, lower bounds, and element access.

\vspace{5pt}
\noindent \textbf{Layer II: Cross-language relations.}
Figure~\ref{fig:proof-hierarchy} shows how the \texttt{Relations} modules connect source and target values and states.
The displayed \texttt{real} constructor of \texttt{RelS} relates a Fortran real value \texttt{v} to a Python value \texttt{w} whenever \texttt{P.toR w = some v}.
In Duo's \texttt{f = 0} example, the Fortran assignment stores a real-valued zero, while Python stores an integer zero.
The constructor relates these values because converting the target integer to its mathematical real value yields the same zero.

The \texttt{RelA} structure relates arrays through its \texttt{shape} and \texttt{elems} fields.
The former requires the Python shape to equal the Fortran extents.
The latter requires \texttt{RelS} for every in-bounds source element and its target counterpart at \texttt{rebase a.lbs idx}.
In Duo's one-based \texttt{parameters} array, this rebases the Fortran subscript \texttt{k+3} to the Python index \texttt{(k+3)-1}, matching the access shown in the code comment.

\texttt{RelState} quantifies over each name \texttt{x} declared in $\Gamma$.
Its existential witnesses \texttt{v} and \texttt{w} require both states to bind that name to values satisfying \texttt{RelV}, which selects the scalar or array relation.
In Duo's example, these bindings include the scalars \texttt{f}, \texttt{z}, and \texttt{k}, together with the array \texttt{parameters}.
Statement soundness obligations use this relation to compare the initial and resulting states.

\vspace{5pt}
\noindent \textbf{Layer III: Modeled translation and contracts.}
The \texttt{Translation} module defines \texttt{Tr.trans}, which recursively maps a source AST to a target AST for the audited fragment.
For Duo's real-valued \texttt{z} and integer \texttt{k}, its \texttt{gfortran} power branch emits a call to the modeled \texttt{f\_powi} helper, corresponding to \texttt{\_f\_powi(z,k)}.
The \texttt{Rules} infrastructure states the preservation contract for such a source/target pair.

Figure~\ref{fig:proof-layer-examples} shows the definition of \texttt{SoundScalar}.
Its parameters \texttt{o}, $\Gamma$, \texttt{e}, and \texttt{e'} fix the semantic oracles, source declarations, and source and target expressions.
The quantifiers range over initial states $\sigma$ and $\tau$ satisfying \texttt{RelState}.
The additional premise \texttt{Conforms o.F} $\Gamma$ $\sigma$ requires the source bindings to have the declared types, array ranks, and lower bounds.
If \texttt{F.eval o.F} $\sigma$ \texttt{e} returns \texttt{some (.s v)}, the contract requires a target scalar \texttt{w} such that \texttt{P.eval o.P} $\tau$ \texttt{e'} returns \texttt{some (.s w)} and \texttt{RelS v w} holds.
Here, \texttt{.s} tags a scalar result.

In Duo's power expression example, \texttt{e} represents \texttt{z**k}, and \texttt{e'} represents the helper call \texttt{\_f\_powi(z,k)}, as indicated in the code comment.
The contract requires a related target result whenever the source power has a defined value in conforming, related states.
Layer IV supplies the proof using sound operand translations and the real-base and integer-exponent premises.

\texttt{SoundPair} extends the contract to scalar or array results through \texttt{RelV}, while \texttt{StmtSound} relates terminating executions and their final states.
\texttt{Refuted} names the negation of \texttt{SoundPair} for counterexample proofs.

\begingroup
\definecolor{proofHBlue}{HTML}{416D95}
\definecolor{proofHGreen}{HTML}{397D70}
\definecolor{proofHOrange}{HTML}{C77B23}
\definecolor{proofHBorder}{HTML}{A8B7C5}
\definecolor{proofHFill}{HTML}{F0F4F8}
\definecolor{proofHFamily}{HTML}{EDF6F2}

\normalsize
\newlength{\proofExampleTwoRefWidth}
\settowidth{\proofExampleTwoRefWidth}{\textbf{\texttt{np.copysign}}}
\setlength{\proofExampleTwoRefWidth}{\dimexpr5\proofExampleTwoRefWidth\relax}
\lstset{style=styleD,language=LeanD,numbers=none,
  xleftmargin=0pt,xrightmargin=0pt,
  rulecolor=\color{black},framerule=\heavyrulewidth}
\newsavebox{\proofLayerThreeBox}
\begin{lrbox}{\proofLayerThreeBox}
\begin{minipage}[t]{\proofExampleTwoRefWidth}
\raggedright\small
\hrule height\heavyrulewidth
\vspace{8pt}
\textbf{\textcolor{proofHBlue}{III}: Preservation contract}
\begin{lstlisting}[frame=b]
-- Duo R3: e = z**k; e' = _f_powi(z,k)
def SoundScalar (o : Oracles) ((*@$\Gamma$@*) : F.Env)
    (e : F.Expr) (e' : P.Expr) : Prop :=
  ∀ (*@$\sigma$@*) (*@$\tau$@*), RelState (*@$\Gamma$@*) (*@$\sigma$@*) (*@$\tau$@*) →
    Conforms o.F (*@$\Gamma$@*) (*@$\sigma$@*) →
    ∀ v, F.eval o.F (*@$\sigma$@*) e = some (.s v) →
      ∃ w, P.eval o.P (*@$\tau$@*) e' = some (.s w) ∧
        RelS v w
\end{lstlisting}
\end{minipage}
\end{lrbox}
\newsavebox{\proofLayerFourBox}
\begin{lrbox}{\proofLayerFourBox}
\begin{minipage}[t]{\proofExampleTwoRefWidth}
\raggedright\small
\textbf{\textcolor{proofHBlue}{IV}: Integer-power rule theorem}
\begin{lstlisting}[frame=b]
-- Duo R3: z**k -> _f_powi(z,k)
example (o : Oracles) ((*@$\Gamma$@*) : F.Env)
    (z k : F.Expr) (z' k' : P.Expr)
    (hz : SoundScalar o (*@$\Gamma$@*) z z')
    (hk : SoundScalar o (*@$\Gamma$@*) k k')
    (hr : RealTyped o (*@$\Gamma$@*) z) (hi : IntTyped o (*@$\Gamma$@*) k) :
    SoundScalar o (*@$\Gamma$@*) (.bin .pow z k)
      (Tr.call2 .f_powi z' k') := by
  exact FNP_R0026_sound o (*@$\Gamma$@*) z k z' k'
    hz hk hr hi
\end{lstlisting}
\end{minipage}
\end{lrbox}
\newsavebox{\proofLayerFiveBox}
\begin{lrbox}{\proofLayerFiveBox}
\begin{minipage}[t]{\proofExampleTwoRefWidth}
\raggedright\small
\hrule height\heavyrulewidth
\vspace{8pt}
\textbf{\textcolor{proofHBlue}{V}: Expression composition theorem}
\begin{lstlisting}[frame=b]
-- Duo R6: e = k+3; e' = (k+3)
example (o : Oracles) ((*@$\Gamma$@*) : F.Env)
    ((*@$\Lambda$@*) : Tr.BoundEnv) (e : F.Expr) (e' : P.Expr)
    (hBounds : BoundEnvOK (*@$\Gamma$@*) (*@$\Lambda$@*))
    (hWF : WFE (*@$\Gamma$@*) e)
    (hTrans : Tr.trans Tr.ifx (*@$\Gamma$@*) (*@$\Lambda$@*) [] e = some e') :
    SoundScalar o (*@$\Gamma$@*) e e' := by
  exact trans_sound o (*@$\Gamma$@*) (*@$\Lambda$@*) [] hBounds
    e hWF e' hTrans
\end{lstlisting}
\end{minipage}
\end{lrbox}
\newsavebox{\proofLayerSixBox}
\begin{lrbox}{\proofLayerSixBox}
\begin{minipage}[t]{\proofExampleTwoRefWidth}
\raggedright\small
\textbf{\textcolor{proofHBlue}{VI}: Whole-kernel theorems}
\begin{lstlisting}[frame=b]
-- SAXPY: y(i) = a*x(i) + y(i), i = 1,...,n
theorem saxpy_trans :
    transStmt (*@$\Gamma$@*)k (*@$\Lambda$@*)k [] saxpyF = some saxpyP := by
  rfl

theorem saxpy_sound (o : Oracles) :
    StmtSound o (*@$\Gamma$@*)k saxpyF saxpyP :=
  transStmt_sound o (*@$\Gamma$@*)k (*@$\Lambda$@*)k [] (*@$\Gamma$@*)k_bounds
    (*@$\Gamma$@*)k_static saxpyF saxpy_wfs saxpyP saxpy_trans
\end{lstlisting}
\end{minipage}
\end{lrbox}

\newlength{\proofThreeFourHeight}
\newlength{\proofFiveSixHeight}
\setlength{\proofThreeFourHeight}{\dimexpr
  \ht\proofLayerThreeBox+\dp\proofLayerThreeBox+
  \ht\proofLayerFourBox+\dp\proofLayerFourBox\relax}
\setlength{\proofFiveSixHeight}{\dimexpr
  \ht\proofLayerFiveBox+\dp\proofLayerFiveBox+
  \ht\proofLayerSixBox+\dp\proofLayerSixBox\relax}
\newlength{\proofThreeFourGap}
\newlength{\proofFiveSixGap}
\setlength{\proofThreeFourGap}{3pt}
\setlength{\proofFiveSixGap}{3pt}
\ifdim\proofThreeFourHeight<\proofFiveSixHeight
  \addtolength{\proofThreeFourGap}{\dimexpr
    \proofFiveSixHeight-\proofThreeFourHeight\relax}
\else
  \addtolength{\proofFiveSixGap}{\dimexpr
    \proofThreeFourHeight-\proofFiveSixHeight\relax}
\fi
\newsavebox{\proofLayersThreeFourBox}
\sbox{\proofLayersThreeFourBox}{\vbox{\offinterlineskip
  \hbox{\usebox{\proofLayerThreeBox}}
  \vskip\proofThreeFourGap
  \hbox{\usebox{\proofLayerFourBox}}}}
\newsavebox{\proofLayersFiveSixBox}
\sbox{\proofLayersFiveSixBox}{\vbox{\offinterlineskip
  \hbox{\usebox{\proofLayerFiveBox}}
  \vskip\proofFiveSixGap
  \hbox{\usebox{\proofLayerSixBox}}}}

\begin{figure}[!tp]
\centering
\begin{minipage}[t]{0.49\linewidth}
\vspace{0pt}\centering
\small\textbf{Examples by proof layer}\par\vspace{8pt}
\nointerlineskip
\resizebox{\linewidth}{!}{\usebox{\proofLayersThreeFourBox}}
\end{minipage}\hfill
\begin{minipage}[t]{0.49\linewidth}
\vspace{0pt}\centering
\small\textbf{Examples by proof layer}\par\vspace{8pt}
\nointerlineskip
\resizebox{\linewidth}{!}{\usebox{\proofLayersFiveSixBox}}
\end{minipage}

\caption{Representative contracts and theorems for Layers III--VI of the proof hierarchy in Figure~\ref{fig:proof-hierarchy}. Layers III--IV (left) show the \texttt{SoundScalar} contract and local theorem for Duo's integer power (R3). Layer V (right, top) uses \texttt{k+3} (R6) under the modeled \texttt{ifx} profile, while Layer VI (right, bottom) shows the translation and soundness proofs for SAXPY. R3 and R6 refer to Table~\ref{tab:rule-composition-mapping} and the example in Figure~\ref{fig:rule-composition}. Snippets are condensed. Preservation claims concern mathematical values under the modeled semantics.}
\Description{The left panel arranges Layers III and IV from top to bottom, showing the scalar preservation contract and the integer-power rule theorem for Duo. The right panel arranges Layers V and VI from top to bottom, showing the expression composition theorem for Duo addition and the translation and soundness proofs for the SAXPY kernel.}
\label{fig:proof-layer-examples}
\end{figure}
\endgroup

\vspace{5pt}
\noindent \textbf{Layer IV: Rule-family theorems.}
The rule-family modules discharge Layer III's contracts for arithmetic, logic, intrinsics, indexing, sections, stores, control flow, and procedures.
Figure~\ref{fig:proof-layer-examples} illustrates the integer-power theorem \texttt{FNP\_R0026\_sound} using Duo's \texttt{z**k} translation (R3) as an example.
The displayed \texttt{example} writes out its four premises and conclusion.
Here, \texttt{z} and \texttt{k} denote source operand ASTs, while \texttt{z'} and \texttt{k'} denote their translated ASTs.
The premises \texttt{hz} and \texttt{hk} require sound translations of the two operands, so each defined source operand evaluation yields a related target value.
The premises \texttt{hr} and \texttt{hi} require a real-valued source base and an integer-valued source exponent.
The conclusion instantiates \texttt{SoundScalar} for the source power node \texttt{.bin .pow z k} and the target helper call \texttt{Tr.call2 .f\_powi z' k'}.
The proof line supplies the four premises to \texttt{FNP\_R0026\_sound}.

The underlying theorem first unfolds the source binary-expression evaluation and extracts the two operand values.
It uses \texttt{hz} and \texttt{hk} to obtain related target operands, then uses the typing premises to identify the source operation as a real base raised to an integer power.
After unfolding target helper-call evaluation, the lemmas \texttt{helper\_f\_powi\_float} and \texttt{helper\_f\_powi\_int} cover the Python base representations permitted by \texttt{RelS}.
The proof considers zero, negative, and positive exponents and establishes \texttt{RelS} between the results for each defined source case.
This gives Duo's power node the same preservation interface as its translated operands.

Generic lemmas such as \texttt{bin\_sound\_of\_rel}, \texttt{call1\_sound\_of\_rel}, \texttt{lift1\_rel}, \texttt{lift2\_rel}, and \texttt{element\_sound} factor out recurring evaluation, lifting, and indexing arguments across other rule families.
Most rule theorems are proved directly from the semantics, relations, and these lemmas.
A theorem can derive from a more basic rule theorem, while unrelated rules can remain siblings that reuse the same generic lemma.
This organization confines a failed proof to the translation decision whose guard, emitted operation, or semantic premise is insufficient.

\vspace{5pt}
\noindent \textbf{Layer V: Composition theorems.}
The expression theorem \texttt{trans\_sound} composes the local contracts by structural recursion over a source AST whose modeled translation succeeds.
Figure~\ref{fig:proof-layer-examples} shows an application with its three premises written explicitly.
\texttt{hBounds} requires the rendered-bound environment $\Lambda$ to supply the same number of bound entries as each declaration in $\Gamma$.
\texttt{hWF} requires \texttt{e} to satisfy the supported syntax and typing conditions in \texttt{WFE}.
\texttt{hTrans} records that \texttt{Tr.trans} returns \texttt{some e'} under the modeled \texttt{ifx} profile.
The conclusion is the \texttt{SoundScalar} contract from Layer III for the complete source/target expression pair.

In Duo's index expression example, \texttt{e} represents \texttt{k+3}, and \texttt{e'} represents the emitted \texttt{(k+3)}.
The well-formedness premise records that \texttt{k} is a declared scalar with numeric type and that the integer literal \texttt{3} is a supported operand.
The recursive call for \texttt{k} applies the name rule \texttt{FNP\_R0010\_sound}, while the call for \texttt{3} applies \texttt{FNP\_R0001\_sound}.
The addition case then passes the two resulting operand contracts to \texttt{FNP\_R0016\_sound}, obtaining soundness of the enclosing \texttt{.bin .add} node.
The displayed proof line supplies the frontend and translation premises to \texttt{trans\_sound}, which performs this composition recursively.
Across 34 local expression rules, each emitter case selects its matching theorem and supplies the recursively established operand contracts.
The local theorems remain independent because their premises quantify over sound operand translations, while the composition theorem connects their applications through the source AST.

The statement theorem \texttt{transStmt\_sound} similarly composes seven statement families and threads \texttt{RelState} through assignment, sequencing, conditionals, supported loops, and state updates.
It also relates Fortran \texttt{EXIT} and \texttt{CYCLE} to target \texttt{break} and \texttt{continue}.
The result is a forward simulation of terminating executions for the supported fragment.

\vspace{5pt}
\noindent \textbf{Layer VI: Whole-kernel theorems.}
Kernel theorems fix the source and target ASTs and frontend environments for a complete translated fragment.
Figure~\ref{fig:proof-layer-examples} shows the translation and soundness proofs for SAXPY, whose loop updates \texttt{y(i) = a*x(i) + y(i)} for \texttt{i = 1,...,n}.
Here, \texttt{saxpyF} and \texttt{saxpyP} encode the complete source and target loop statements.
The declaration environment \texttt{$\Gamma$k} assigns integer types to \texttt{n} and \texttt{i}, a real type to \texttt{a}, and one-dimensional real-array types with lower bound one to \texttt{x} and \texttt{y}.
The rendered-bound environment \texttt{$\Lambda$k} supplies the corresponding target lower-bound expressions.
The target AST uses \texttt{range(1,n+1)} and indices \texttt{i-1}, and includes loop-completion code to preserve the final value of \texttt{i}.

The theorem \texttt{saxpy\_trans} states that \texttt{transStmt} returns \texttt{some saxpyP} for \texttt{saxpyF} in these environments.
Its \texttt{rfl} proof checks this equality by unfolding the concrete definitions.
The theorem \texttt{saxpy\_sound} then applies Layer V's \texttt{transStmt\_sound}.
The proof line passes \texttt{$\Gamma$k\_bounds}, which establishes \texttt{BoundEnvOK}, and \texttt{$\Gamma$k\_static}, which establishes that the declared lower bounds are integer literals.
It also passes \texttt{saxpy\_wfs}, the well-formedness proof for the loop and its body, and \texttt{saxpy\_trans}, the translation equality.
These facts discharge the frontend and translation premises of the composition theorem for this particular kernel.

The conclusion \texttt{StmtSound o} \texttt{$\Gamma$k saxpyF saxpyP} quantifies over every pair of related initial states whose source state conforms to the declarations.
For each terminating source execution described by \texttt{F.Exec}, it requires a target execution described by \texttt{P.Exec}, with the control outcome related by \texttt{outMap} and the final states related by \texttt{RelState}.
It also preserves source-state conformance.
The array size \texttt{n} remains symbolic, so the theorem covers every defined terminating execution admitted by the modeled semantics.
The guarantee concerns mathematical values in this fragment, under the premises of \texttt{StmtSound}.

\section{Evaluation}
\label{sec:evaluation}

\subsection{Benchmarks and Research Questions}
\label{sec:eval-setup}

\begin{table}[t]
\centering
\fontsize{7.2}{8.6}\selectfont
\setlength{\tabcolsep}{2.5pt}
\setlength{\arrayrulewidth}{0.4pt}
\arrayrulecolor{black}
\renewcommand{\arraystretch}{1.08}
\caption{Scientific software, evaluation benchmarks, and testing environments.}\label{tab:evaluation-cases}
\begin{tabular}{@{}c|c|l|r||l|c|c@{}}
\specialrule{0.8pt}{0pt}{0pt}
\multicolumn{1}{c}{\textbf{Software}} &
\multicolumn{1}{c}{\textbf{Source}} &
\multicolumn{1}{c}{\textbf{Domain}} &
\multicolumn{1}{c}{\textbf{LoC}} &
\multicolumn{1}{c}{\textbf{Benchmark}} &
\multicolumn{1}{c}{\textbf{Source}} &
\multicolumn{1}{c}{\textbf{Env}} \\
\specialrule{0.4pt}{0pt}{0pt}
CLUBB & \citep{software-clubb} & Atmos. physics & 144,864 & Single-column cases & \citep{bench-clubb} & H1 \\
RRTMG & \citep{software-rrtmg-lw,software-rrtmg-sw} & Radiation & 173,296 & Standard atmospheres & \citep{bench-rrtmg-lw,bench-rrtmg-sw} & H1 \\
CISM & \citep{software-cism} & Land ice & 77,332 & Halfar-HO & \citep{bench-cism} & H1 \\
WW3 & \citep{software-ww3} & Waves & 183,354 & tp1.1 and ts1 & \citep{bench-ww3} & H1 \\
COSP2 & \citep{software-cosp2} & Satellite obs. & 18,411 & UKMO offline test & \citep{bench-cosp2} & H1 \\
CARMA & \citep{software-carma} & Aerosols & 21,705 & Regression tests & \citep{bench-carma} & H1 \\
CVMix & \citep{software-cvmix} & Ocean mixing & 7,607 & Regression tests & \citep{bench-cvmix} & H1 \\
MARBL & \citep{software-marbl} & Biogeochemistry & 19,023 & Regression tests & \citep{bench-marbl} & H1 \\
FATES & \citep{software-fates} & Ecosystems & 50,277 & Leaf-biophysics smoke test & \citep{bench-fates} & H1 \\
MFC & \citep{software-mfc} & Multiphase flow & 44,799 & 1D boundary conditions & \citep{bench-mfc} & H1 \\
QE & \citep{software-qe} & Electronic struct. & 570,099 & XClib; DFT-D3 & \citep{bench-qe-xclib,bench-qe-d3} & H1 \\
RAMSES & \citep{software-ramses} & Astrophysics & 159,286 & Hydrodynamics tests & \citep{bench-ramses} & H1 \\
NJOY2016 & \citep{software-njoy} & Nuclear data & 100,487 & Tests 01, 02, 09, 10 & \citep{bench-njoy} & H1 \\
Duo & \citep{software-duo} & Spectroscopy & 75,074 & PS and SO & \citep{bench-duo} & H1 \\
\multirow[t]{3}{*}{CAM-SIMA} & \multirow[t]{3}{*}{\citep{software-cam-sima}} & \multirow[t]{3}{*}{Atmos. physics} & \multirow[t]{3}{*}{137,829} & Kessler & \citep{bench-cam-kessler} & H2 \\
 &  &  &  & Held--Suarez & \citep{bench-cam-held-suarez} & H2 \\
 &  &  &  & TJ2016 & \citep{bench-cam-tj2016} & H2 \\
\bottomrule
\end{tabular}
\end{table}

We evaluate \sys on 17 benchmarks drawn from 15 real-world software. 
The benchmarks span diverse scientific domains, including electronic structure, multiphase flow, nuclear data, astrophysics, molecular spectroscopy, ocean mixing, biogeochemistry, radiation, satellite observations, aerosols, terrestrial ecosystems, land ice, waves, and atmospheric physics.
These are purposively selected case studies with available and compilable Fortran source and native benchmark drivers. 
Their variety exercises different numerical and state-management requirements. 

Table~\ref{tab:evaluation-cases} summarizes the software and benchmarks. 
For each one, we have the software name (\textbf{Software}), its origin (\textbf{Source}), scientific domain (\textbf{Domain}), the total line of Fortran code (\textbf{LoC}), the corresponding benchmark (\textbf{Benchmark}), and the testing environment (\textbf{Env}).

We conduct the experiments on two high-performance computing
(HPC) systems, referred to as HPC1 and HPC2 to preserve anonymity
and abbreviated as H1 and H2 in Table~\ref{tab:evaluation-cases}.
HPC1 uses AMD EPYC 74F3 CPU nodes.
HPC2 uses dual-socket AMD EPYC 7763 CPU nodes, with 128 cores
and 256\,GB of DDR4 memory per node.
The software environment on HPC1 uses GCC/gfortran 14.2.0,
Python 3.12.14, NumPy 2.5.3, SciPy 1.18.1, and Open MPI 5.0.6.
The software environment on HPC2 uses GCC/gfortran 14.3.0,
Python 3.11.14, NumPy 2.4.6, and Cray MPICH 8.1.32.

To evaluate \sys against these benchmarks, we design the following three research questions:
\begin{itemize}
    \item \textbf{RQ1. Real-world Translation Effectiveness:} How many the benchmark computation can \sys translate to achieve bitwise-identical results?
    \item \textbf{RQ2. Rule Assurance:} Can the Auditor identify unsound or overly general translation rules despite agreement on the observed application tests?
    \item \textbf{RQ3. Rule Reusability:} To what extent can executable rules learned and accumulated be reused across real-world scientific software, and can reusing rules compensate model calls and token usage for rule generation and token generation?
\end{itemize}

\subsection{RQ1. Real-world Translation Effectiveness}
\label{sec:eval-rq1}

\noindent \textbf{Setup and baseline.}
We use Claude Opus~5.5 as the direct LLM baseline, because it was the strongest model available to us when we conducted the experiments and submitted this paper.
We construct this baseline from \sys by removing the Auditor, Human-Expert, and reusable translation rules, while retaining the minimal core's feedback loop.
Opus~5.5 generates Python code directly, and the Executor compares its behavior with native Fortran and returns execution failures and numerical mismatches to the model.
The model uses this feedback to revise the generated code. 
We allow repeated repair attempts without imposing a token budget, continuing until the generated code passes the execution checks or the model decides that it cannot resolve the remaining failures.

We check output in scalar points for bitwise-identical agreement.
One scalar point is one unmasked scalar output value recorded through instrumentation to compare with its native Fortran reference for a given test input.
A scalar output contributes one point, and an array output contributes one point per compared element.
For example, a $10\times20$ output array contributes 200 points per test input.
It should be noted that passing tests establishes agreement on the exercised inputs, but does not establish semantic equivalence on untested inputs or execution paths.

\newsavebox{\translationeffectivenesstable}
\begin{table}[t]
\centering
\fontsize{7.2}{8.6}\selectfont
\setlength{\tabcolsep}{2.5pt}
\setlength{\arrayrulewidth}{0.4pt}
\arrayrulecolor{black}
\renewcommand{\arraystretch}{1.08}
\caption{Translation coverage and bitwise agreement on the benchmark scope.
$T/R$ denotes translated/reached computational subprograms. Scope reports source lines of code (SLOC) in the reached subprograms.
Tests reports comparison scale in scalar points: each unmasked scalar output value contributes one point, and an array contributes one point per compared element.
\ding{51} indicates that all scored tests pass. In the Opus~5.5 column, \ding{55}~$n$ denotes $n$ failing routine calls.}
\label{tab:translation-effectiveness}
\begin{lrbox}{\translationeffectivenesstable}
\begin{tabular}{@{}l|c|r|r||c|c@{}}
\specialrule{0.8pt}{0pt}{0pt}
\multicolumn{1}{c}{\multirow{2}{*}{\textbf{Benchmark}}} &
\multicolumn{1}{c}{\multirow{2}{*}{\boldmath$T/R$}} &
\multicolumn{1}{c}{\multirow{2}{*}{\shortstack{\textbf{Scope}\\\textbf{(SLOC)}}}} &
\multicolumn{1}{c}{\multirow{2}{*}{\shortstack{\textbf{Tests}\\\textbf{(scalar points)}}}} &
\multicolumn{2}{c}{\textbf{Bitwise identical}} \\
\cline{5-6}
\noalign{\vskip 2pt}
\multicolumn{1}{c}{} & \multicolumn{1}{c}{} & \multicolumn{1}{c}{} & \multicolumn{1}{c}{} &
\multicolumn{1}{c}{\textbf{Opus 5.5}} & \multicolumn{1}{c}{\textbf{\sys}} \\
\specialrule{0.4pt}{0pt}{0pt}
CLUBB: Single-column cases & 169/169 & 22,606 & 5,122,717 & \ding{51} & \ding{51} \\
RRTMG: Standard atmospheres & 107/107 & 13,320 & 19,649,812 & N/A & \ding{51} \\
CISM: Halfar-HO & 91/92 & 14,747 & 55,081,217 & \ding{51} & \ding{51} \\
WW3: tp1.1 and ts1 & 54/54 & 7,036 & 35,270,521 & \ding{55}~2 & \ding{51} \\
COSP2: UKMO offline test & 74/75 & 9,583 & 141,457,253 & \ding{55}~5 & \ding{51} \\
CARMA: Regression tests & 118/122 & 9,919 & 28,129,723 & N/A & \ding{51} \\
CVMix: Regression tests & 27/27 & 1,870 & 19,853 & \ding{51} & \ding{51} \\
MARBL: Regression tests & 79/79 & 5,431 & 5,118,479 & \ding{51} & \ding{51} \\
FATES: Leaf-biophysics smoke test & 22/22 & 1,002 & 1,299 & \ding{51} & \ding{51} \\
MFC: 1D boundary conditions & 49/49 & 4,212 & 2,302,066 & \ding{51} & \ding{51} \\
QE: XClib; DFT-D3 & 94/94 & 8,242 & 12,084,364 & \ding{51} & \ding{51} \\
RAMSES: Hydrodynamics tests & 19/19 & 1,439 & 10,609,501 & \ding{51} & \ding{51} \\
NJOY2016: Tests 01, 02, 09, 10 & 57/57 & 4,190 & 661,546 & \ding{51} & \ding{51} \\
Duo: PS and SO & 18/18 & 794 & 611,042 & \ding{51} & \ding{51} \\
CAM-SIMA: Kessler & 30/30 & 758 & 2,902,113 & \ding{51} & \ding{51} \\
CAM-SIMA: Held--Suarez & 13/13 & 346 & 1,880,549 & \ding{51} & \ding{51} \\
CAM-SIMA: TJ2016 & 14/14 & 630 & 3,033,528 & \ding{51} & \ding{51} \\
\bottomrule
\end{tabular}
\end{lrbox}
\usebox{\translationeffectivenesstable}

\end{table}

\vspace{5pt}
\noindent \textbf{Results analysis.}
Table~\ref{tab:translation-effectiveness} summarizes computational source scope, translation coverage, and the observed agreement with native Fortran.
Across the 15 systems, \sys translates 1,015 of the 1,021 subprograms reached during benchmark execution into Python (99.4\%).
The source scope spans 105,631 SLOC and includes 37 distinct subprograms across the three CAM-SIMA schemes and 54 across the two WW3 builds.

Six reached routines are not translated.
Four CARMA creation routines require allocation transfer for arrays of records within record arguments.
COSP2's \texttt{cosp\_simulator} performs direct-access I/O on an already connected unit.
In CISM, the subroutine \texttt{glissade\_initialise} accesses nested records and private module state.
They do not prevent benchmark execution.
We retain their Fortran implementations and invoke them through \texttt{f2py} wrappers.
We dump the data passed across these interfaces and make the recorded results available to both the direct LLM baseline and \sys, allowing them to replay these calls without translating the six routines.
These six cases expose limitations in cross-language data transfer and access to application state that require runtime and interface support beyond source-code translation.
As part of future work, we plan to extend \sys by supplying templates that provide explicit packing and unpacking interfaces for nested records and allocation metadata, a Fortran I/O bridge operating on connected unit, and module-local accessor procedures for exposing and synchronizing private state.


\sys achieves bitwise-identical with native Fortran on all evaluated benchmark tests under the recorded execution configurations.
In comparison, direct LLM translation produces seven failing routine calls across two benchmarks and fails to establish reliable numerical agreement on two others, reported as N/A.
The seven failures occur in COSP2 and WW3.
Five occur in COSP2's \texttt{quickbeam\_optics}, and two occur in WW3, one each in \texttt{w3iogomd} and \texttt{w3src3md}.
The diagnosed COSP2 failures and the \texttt{w3iogomd} failure involve reproducing the reference compiler's vector mathematical functions in \texttt{libmvec}.
The \texttt{w3iogomd} failure includes an 18-ULP discrepancy in a fourth-power calculation.
The \texttt{w3src3md} hidden-test failure remains undiagnosed.

The two N/A entries concern CARMA and RRTMG, whose original Fortran programs read allocated state before all relevant elements have been assigned.
Direct LLM translation did not identify these uninitialized reads or distinguish source-program defects from translation defects, and therefore failed to establish reliable numerical agreement.
Our validation identified the initialization defects and confirmed agreement under controlled initialization.
CARMA matches 25 of 27 application tests under ordinary execution.
Six tests are sensitive to uninitialized state, and all six match when both executions zero-fill heap allocations.
Together with the ordinary passing tests, these results cover all 27 tests.
RRTMG matches all 16 shortwave tests and 13 of 14 longwave tests under ordinary initialization, and all 14 longwave tests under zero initialization.

From the results, we can observe that using LLMs to propose translation rules does not limit \sys to the capabilities of the underlying model, because \textbf{Lean proof provides an additional feedback channel}. 
The direct LLM baseline retains failures despite repeated self-correction using execution feedback, whereas our system achieves agreement on the evaluated benchmarks. 
Lean-checked refutations expose semantic defects or missing applicability conditions in LLM-proposed rules and guide their correction beyond what execution feedback alone reveals.
While the motivating example \texttt{ANINT} already illustrates this process, RQ2 (Section~\ref{sec:eval-rq2}) presents additional concrete cases.

\vspace{5pt}
\noindent \textbf{Developer adoption.}
CLUBB developers are reviewing the translated Python code and have already integrated part of it into their codebase.

\subsection{RQ2. Rule Assurance}
\label{sec:eval-rq2}

Evaluating against the 15 scientific software, the Lean proof library in \sys accumulates 42 Lean modules and 190 distinct theorem declarations. 
Building the modules, checking the cited declarations, auditing the namespace, and checking the two supplementary developments takes 114.6\,s in one measured run with precompiled Mathlib dependencies. 
The manual correspondence and assumption audit required
approximately one person-week in total.

We report assurance through 188 audit records which covers all 193 executable rules.
Each record identifies the translation decision or modeled subcase being audited and records its source span, assumptions, and supporting theorems or execution evidence. 
An audit record may cite several theorems, several records may share one, and unresolved or provenance records may cite none.

In general, out of the 188 audit records, 121 records (64.4\%) have checked soundness proofs for claims within the Lean fragment shown in Figure~\ref{fig:assurance-domains}.
There are 25 records (13.3\%) regarding behaviors within the supported translation fragment but beyond the current Lean fragment, requiring model extensions and external contracts.
Compared with direct LLM's blackbox, these records also contribute assurance by exposing candidate defects and making unresolved correctness assumptions explicit.
The remaining 42 records (22.3\%) are unfinished.
When \sys produces a new translation rule and a corresponding soundness obligation, several rounds of refinement against more software is needed to capture the behaviors precisely before they can check a soundness proof or a counterexample.
We provide more detailed discussion below.

\vspace{5pt}
\noindent \textbf{Soundness proofs.}
A rule is sound when, for every source instance admitted by its pattern
and guard, executing the emitted Python produces a value or state effect
corresponding to that of the Fortran construct under the stated semantic and
binding assumptions.
Of the 121 records with soundness proofs, 80 establish this property using only the stated assumptions and 41 establish it under additional applicability premises.
Further, there are 31 established after counterexample-guided refinement. 

For the 41 records with additional applicability premises, the proof applies
only when extra conditions hold. Because the general rule itself does not
guarantee those conditions, the proof makes no claim about inputs that violate
them.
For example, the positive-step array-section proof requires
\texttt{SectionInRange}, which keeps the translated slice edges within the
array. Without this premise, Fortran can denote an empty section such as
\texttt{a(1:-1)}, while Python's negative-index adjustment can select actual
elements. The premise must therefore be enforced by a guard or supplied by an
input contract. For sections, the supplementary state-local formulation makes
the bounds condition explicit for the current array and execution state.

The 31 records with counterexample-guided repairs illustrate how LLM interaction with Lean provides an additional feedback channel that strengthens ForTress beyond the capabilities of the underlying LLM.
The motivating \texttt{ANINT}-to-\texttt{np.round} example is a representative case. 
The following is another case tracing the process from a checked
refutation to a proved repair.

Consider \texttt{value = real(radix ** level, 8)} with
\texttt{(radix, level) = (2, -1)}.
Because both operands are integers, Fortran evaluates
\texttt{radix ** level} using integer exponentiation, producing zero, and
then converts it to \texttt{0.0}.
The earlier rule instead emits \texttt{float(radix ** level)}, for which
Python evaluates \texttt{2 ** -1} as \texttt{0.5}.
Lean checks the refutation proof constructed by the LLM using this counterexample.
The repaired rule checks that both operands are integer-valued and emits
\texttt{np.float64(\_f\_ipow(radix, level))}, where \texttt{\_f\_ipow}
implements Fortran's integer-exponentiation behavior before the conversion
to real.
Under the premises that the translated operands have the correct integer
types and preserve their source values, the Lean theorem proves the repaired
transformation sound for arbitrary runtime integer exponents.
In contrast, the Opus 5.5 baseline insists on \texttt{float(radix ** level)}, leaving the negative-exponent discrepancy unresolved.


\vspace{5pt}
\noindent \textbf{Identified uncertainties.}
The Auditor also provides assurance by exposing candidate defects and identifying where soundness remains unresolved.
These findings highlight differences in how Fortran and Python handle persistent state, data representation, and observable effects, as well as dependencies on operator-supplied behavior.

Five records contain execution counterexamples without separate Lean refutations. 
For example, a Fortran local variable with a declared initializer retains its value across calls through implicit \texttt{SAVE}, whereas a candidate translation that reinitializes it at every entry loses updates from previous calls. 
Paired execution confirms this defect. 

For eleven audit records, the current Lean model lacks the semantic definitions needed to prove the corresponding translation claims.
For example, \texttt{TRANSFER} depends on the bit representation of a value, which the current exact-arithmetic value model does not capture.
Proving its translation sound therefore requires a bit-level model.
In addition, derived types require modeling component state, while I/O requires modeling observable input and output effects.

The nine external-contract records identify guarantees that depend on operator-supplied configuration or extensions. 
An intrinsic override, for example, can replace built-in emission. 
Its preserved values and state effects must be established before the replacement can be admitted for use.
The audit exposes these dependencies as explicit contracts that must hold for the translation to preserve source behavior.



\vspace{5pt}
\noindent \textbf{Unfinished.}
The 42 unfinished records comprise 34 specified claims without a completed proof or refutation, seven records without separate value-level obligations, and one open array-lifting claim.
Completing these records requires proving or refuting the specified claims,
formulating the missing obligations, and lifting scalar soundness results to
conforming arrays and scalar-expansion cases.
As \sys translates more software, additional constructs and usage contexts
can expose missing definitions or assumptions and guide this refinement,
making them to reach checked results.

\subsection{RQ3. Rule Reusability}
\label{sec:eval-rq3}

\noindent \textbf{Measurement Method.}
To measure to what extend can executable rules learned and accumulated be reused across real-world scientific software, we adopt the train--test separation commonly used to evaluate machine learning models.
Software from which \sys acquires translation rules forms the ``training set'', while one held-out software system forms the ``test set'' for measuring rule reuse. 
We repeat this procedure with each of the 15 software systems as the test set.
To record the number of rule application, we instrument the translator without comprising the translation consistency.
Repeated translation attempts at the same source site count as separate application events but contribute only once to the distinct source-site count.


To answer the question whether reusing rules compensate model calls, token usage, and translation cost for rule generation and token generation. we run \sys end to end on all 15 software systems. 
In a separate environment, we run the same software using the direct LLM baseline from RQ1 (Section~\ref{sec:eval-rq1}), removing the Auditor, Human-Expert, and reusable translation rules while retaining the minimal core's feedback loop. 
Both workflows continue using Claude Opus~5.5.

We estimate LLM costs from client-recorded token usage and the prices
applicable to each run, rather than subscription payments.
We record input, output, cache-write, and cache-read tokens separately,
including thinking tokens in the output count.
We deduplicate usage-bearing assistant messages by identifier across
transcript blocks. Request counts measure model calls, and each translation
or repair attempt may involve multiple calls.
We account separately for development, direct translation, fixture
construction, and evidence collection.

\begin{figure}[t]
\centering
\def\rqReuseNumerators{{
  0,34,46,48,53,53,53,53,53,53,53,53,53,53,53, 
  34,34,38,40,42,43,43,43,43,43,43,43,43,43,43, 
  45,49,49,63,65,65,65,66,66,66,66,66,66,66,66, 
  44,48,62,62,68,71,71,71,71,71,71,71,71,71,71, 
  48,54,68,74,74,78,78,79,79,80,81,81,82,82,82, 
  40,45,53,56,60,60,60,60,60,60,60,60,63,63,63, 
  37,39,42,43,43,43,43,43,43,43,43,43,43,43,43, 
  39,42,46,48,49,49,49,49,49,50,50,50,50,50,50, 
  30,30,31,33,33,33,33,33,33,33,33,33,33,33,33, 
  39,41,46,47,48,48,48,49,49,49,49,49,49,49,49, 
  37,39,46,47,48,48,48,48,48,48,48,48,48,48,49, 
  29,30,33,34,34,34,34,34,34,34,34,34,34,34,34, 
  33,37,41,44,46,49,49,49,49,49,49,49,49,49,49, 
  33,36,45,46,47,47,47,47,47,47,47,47,47,47,47, 
  34,34,38,38,38,39,39,39,39,39,40,40,40,40,40 
}}
\def\rqReuseDenominators{{56,45,72,74,92,69,43,51,33,49,49,34,50,49,40}}
\definecolor{rqReuseLow}{HTML}{F4A6A6}
\definecolor{rqReuseMid}{HTML}{FEE08B}
\definecolor{rqReuseHigh}{HTML}{006837}
\def\rqReuseSetColor#1{%
  \ifdim#1pt<50pt
    \pgfmathsetmacro{\rqReuseBlend}{2*#1}%
    \colorlet{rqReuseCellColor}{rqReuseMid!\rqReuseBlend!rqReuseLow}%
  \else
    \pgfmathsetmacro{\rqReuseBlend}{2*(#1-50)}%
    \colorlet{rqReuseCellColor}{rqReuseHigh!\rqReuseBlend!rqReuseMid}%
  \fi
}
\cachetikzfigure{rule-reuse}{figs/hot.tex}
\begin{tikzpicture}[x=0.60cm,y=-0.35cm,
  every node/.style={font=\fontsize{7.2}{8.6}\selectfont,inner sep=1pt}]
  \node[font=\fontsize{8}{9.6}\selectfont\bfseries] at (7.5,-5.40)
    {Cumulative rule coverage};
  \node at (7.5,-4.46) {Rule sources accumulated from left to right};
  \draw[->,line width=0.5pt] (0,-3.86) -- (15,-3.86);
  \node[rotate=90] at (-2.75,7.5) {Target software};
  \foreach \software [count=\idx] in
    {CLUBB,RRTMG,CISM,WW3,COSP2,CARMA,CVMix,MARBL,FATES,MFC,QE,RAMSES,NJOY2016,Duo,CAM-SIMA} {
    \node[anchor=east] at (-0.18,\idx-0.5) {\software};
    \node[anchor=west,rotate=65] at (\idx-0.5,-0.17) {\software};
  }
  \foreach \testrow in {1,...,15} {
    \foreach \sourcecol in {1,...,15} {
      \ifnum\testrow=\sourcecol
        \fill[black!12] (\sourcecol-1,\testrow-1) rectangle (\sourcecol,\testrow);
        \node[text=black!55] at (\sourcecol-0.5,\testrow-0.5) {\textemdash};
      \else
        \pgfmathsetmacro{\covered}{\rqReuseNumerators[(\testrow-1)*15+\sourcecol-1]}
        \pgfmathsetmacro{\denominator}{\rqReuseDenominators[\testrow-1]}
        \pgfmathsetmacro{\shade}{100*\covered/\denominator}
        \pgfmathtruncatemacro{\celllabel}{round(\shade)}
        \rqReuseSetColor{\shade}
        \fill[rqReuseCellColor]
          (\sourcecol-1,\testrow-1) rectangle (\sourcecol,\testrow);
        \ifnum\celllabel>88
          \node[text=white] at (\sourcecol-0.5,\testrow-0.5) {\celllabel};
        \else
          \node[text=black] at (\sourcecol-0.5,\testrow-0.5) {\celllabel};
        \fi
      \fi
    }
  }
  \draw[step=1,white,line width=0.3pt] (0,0) grid (15,15);
  \draw[black!70,line width=0.4pt] (0,0) rectangle (15,15);
  \begin{scope}[xshift=0.5cm]
  \node at (15.8,2.2) {Reuse (\%)};
  \foreach \step in {0,...,49} {
    \pgfmathtruncatemacro{\legendshade}{100*\step/49}
    \rqReuseSetColor{\legendshade}
    \fill[rqReuseCellColor]
      (15.55,{12-0.18*\step}) rectangle (16.0,{12-0.18*(\step+1)});
  }
  \draw[black!60,line width=0.3pt] (15.55,3) rectangle (16.0,12);
  \node[anchor=west] at (16.12,3.1) {100};
  \node[anchor=west] at (16.12,11.9) {0};
  \filldraw[fill=black!12,draw=white,line width=0.3pt]
    (15.55,14.0) rectangle (16.0,14.6);
  \node[anchor=west] at (16.12,14.3) {Excluded};
  \end{scope}
\end{tikzpicture}
\caption{Cumulative coverage of observed translation rules across 15 software
systems. At column $j$, the accumulated pool
is the union of rules observed in columns 1 through $j$, excluding the
test software. Each number is the rounded percentage of the test
software's observed rule types covered by this pool. The denominator is
the number of rule types observed in that software, rather than all 193
rules. Thus, 100\% means that all of its observed rule types also occur
in the donor pool. Gray diagonal cells mark self-exclusion.} 
\Description{A measured 15 by 15 heatmap. Rows identify target software and
columns identify software whose observed rules accumulate from left to
right. Numeric labels and a light-red--yellow--green scale show percentage coverage
from zero to one hundred; gray diagonal cells exclude the target from its
own donor pool. Final coverage by other systems ranges from 89.1 to 100
percent. The matrix measures retrospective rule overlap.}
\label{fig:cumulative-rule-reuse}
\end{figure}
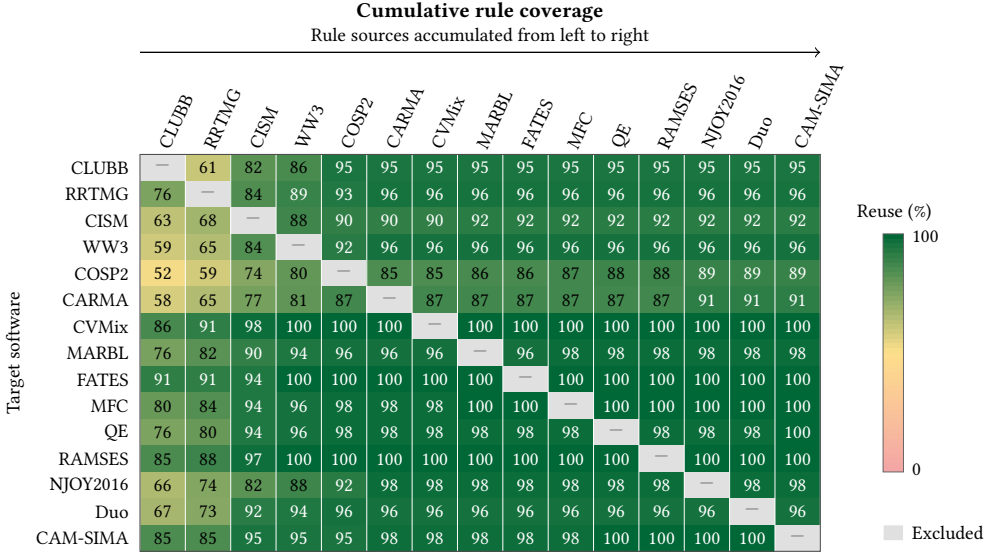

\vspace{5pt}
\noindent \textbf{Rule reuse across software.}
Figure~\ref{fig:cumulative-rule-reuse} shows the extent to which executable rules accumulated from the ``training set'' software can be reused in the ``test set'' software.
Rows identify test software, and columns identify rule-source software in a fixed order.

In Figure~\ref{fig:cumulative-rule-reuse}, reuse rates are monotonically nondecreasing because each additional software can only add rule types to the accumulated pool.
By the final column, 14 of the 15 software exceed 90\% reuse, and six reach 100\%. Even with CLUBB as the sole rule source, the lowest observed reuse
rate is 52\%, showing that rules accumulated from a single software could already cover a substantial share of other systems' translation needs.
Several systems reach full coverage before all donors are included.
FATES, CVMix, and RAMSES reach 100\% after the first four rule-source systems, through WW3. MFC reaches 100\% at the MARBL column, after the first eight rule-source systems.

Individual software can contribute markedly different gains.
Adding CISM raises CLUBB's reuse rate from 61\% to 82\%, adding WW3
raises CISM's from 68\% to 88\%, and adding CISM raises WW3's from
65\% to 84\%. These jumps occur when a new donor supplies rule types
used by the target but absent from the preceding pool. 
They suggest that
similarities in language-feature use make some software more effective
rule sources for particular targets.
Figure~\ref{fig:single-source-rule-reuse} in Appendix~\ref{app:single-source-rule-reuse} shows how coverage varies across individual rule sources, confirms this trend, and further illustrates that, for some software, a single source is sufficient to cover the entire observed rule set.

COSP2 is the clearest exception to the high final coverage.
It starts at the lowest reuse rate, 52\%, and remains the only software below 90\% at the end, reaching 89\%, suggesting that some of its Fortran feature usage is more distinctive.
CARMA shares similar situation.
Other cases also exhibit plateaus. 
For example, Duo reaches 96\% at the COSP2 column and gains no further coverage from the remaining donors.
Accumulating rules from more software with complementary language-feature use is therefore needed to extend coverage.

The three most frequently reused rules handle recurring expression operations.
\texttt{R-name.reference} translates source names into Python references.
\texttt{R-index.shift\_unit} rebases array subscripts with unit lower bounds
to zero-based Python indices, mapping \texttt{i} to \texttt{i-1}.
\texttt{R-arithmetic.scalar\_fallback} combines translated operands with
ordinary scalar operators, such as addition and multiplication.
Their prevalence reflects how often these basic operations recur within
larger expressions across software. 
Once learned, these rules can be applied and composed at many source sites.
Table~\ref{tab:rule-reuse-counts} in Appendix~\ref{app:rule-reuse-frequency}
reports per-software counts for the ten most frequently observed rules.

\begin{figure}[t]
\centering
\includegraphics[width=0.65\linewidth]{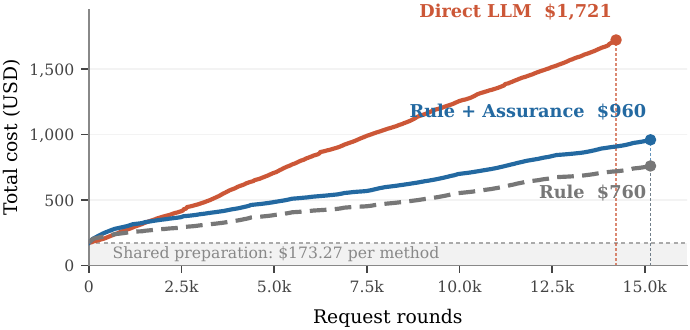}
\caption{Cumulative estimated LLM usage cost over request rounds for
Direct LLM, Rule+Assurance, and Rule-only (estimated).
Each round is one recorded model request. The shaded band marks the
shared preparation cost included in each curve.}
\Description{Three curves plot cumulative estimated LLM usage cost against
recorded request rounds. All begin at a shared preparation cost of 173.27
dollars. The solid orange Direct LLM curve ends at approximately 1,721 dollars,
the dashed gray Rule curve at 760 dollars, and the solid blue Rule+Assurance
curve at 960 dollars. Both rule-based settings use more request rounds
than Direct LLM while ending at lower costs.}
\label{fig:cost-rounds}
\end{figure}

\vspace{5pt}
\noindent \textbf{Token Cost.}
\sys adds reusable rule generation and Lean proof to the direct LLM baseline's minimal core. 
With this additional determinism and assurance, its LLM token cost remains lower than the direct LLM. 
Figure~\ref{fig:cost-rounds} shows token cost against model requests for three settings.
These settings are Direct LLM using Opus~5.5, full \sys (Rule+Assurance), and a Rule-only \sys without assurance.
Because executable rules and Lean assurance are integrated in \sys, we estimate the Rule-only cost by excluding requests issued by the Auditor.

All three start from a common \$173.27 recording and reference-preparation cost.
As model requests accumulate, costs increase in all three settings, following an approximately linear trend at different rates.
Direct LLM finally reaches \$1,721 to translate the 15 scientific software, compared with \$960 for \sys and \$760 for Rule-only \sys.
Adding reusable rules reduces the recorded cost by 55.8\%, and including assurance still leaves \sys 44.2\% below Direct LLM.

Rule-only and Rule+Assurance initially cost slightly more than Direct LLM.
Developing reusable rules requires identifying source patterns,
implementing emission logic, and validating applicability, while assurance
adds semantic definitions, proof obligations, and proof generation.
These activities incur costs before the resulting rules are reused across
software and start offsetting the cost.

From Figure~\ref{fig:cost-rounds}, we can also observe that Rule-only and Rule+Assurance use more request rounds than Direct LLM but still have lower costs.
This is because their requests are of different types.
Direct LLM translates and repairs files and constructs validation fixtures and reference outputs.
Rule-only develops reusable translation rules, while Rule+Assurance additionally develops semantic definitions and proofs.
Requests also contain different proportions of uncached input, output (including thinking), cache reads, and cache writes, which are priced differently.
Within the accounted scope, Direct LLM produces 25.88 million output tokens and writes 65.63 million tokens to the prompt cache, compared with 8.95 million and 21.04 million for Rule-only, and 11.36 million and 32.53 million for Rule+Assurance.
Both rule-based settings thus use fewer of the more expensive output and cache-write tokens.
Instead, lower-priced cache reads account for 98.7\% of Rule-only's tokens and 98.4\% of Rule+Assurance's, compared with 97.4\% for Direct LLM.

\section{Threats to Validity}
\label{sec:threats}
Though our evaluation contains 15 scientific software across different domains, their results may not generalize to unexecuted branches, longer trajectories, or other language constructs. 
All evaluated systems supplied development examples, limiting conclusions about transfer to unseen software.

Felleisen's notion of macro-expressibility makes expressibility relative to restrictions on behavior-preserving translations~\cite{felleisen1991expressive}.
Our successful translations do not establish that reusable rules can cover all Fortran constructs and behaviors.
For a fixed rule model, target runtime, and preservation criterion, the permitted rule compositions impose an upper bound on achievable coverage.
Our rule library continues to expand, but we have not characterized this bound or quantified the gap between current and maximally achievable coverage.
We also have no completeness guarantee that our generation and refinement process can reach that bound.
The six routines retained in Fortran in Section~\ref{sec:eval-rq1} illustrate limitations in the current system's support for cross-language data transfer and access to application state.
We plan to address these cases with targeted runtime and interface templates, whose effectiveness and generality remain to be established.

Output normalization and controlled initialization restrict the behaviors
compared by the tests.
The observed bitwise-identical holds under the recorded compiler and runtime versions, optimization settings, mathematical libraries, and hardware configurations.
Different environments may change operation lowering and floating-point rounding, as illustrated by the \texttt{libmvec} and exponentiation cases.
Numerical agreement therefore requires renewed validation when these configurations change.
Different task boundaries, omitted development costs, and the absence of an
application-level LLM repair comparison prevent estimating total migration
savings. Finally, formal results
depend on the modeled semantics and their correspondence with production
rules. Exact-arithmetic proofs alone do not establish IEEE floating-point
equivalence.

\section{Related Work}
\label{sec:related}
\noindent \textbf{LLM-driven code translation.}
Recent neural and large-language-model approaches have
substantially expanded the range of translations that can be automated.
TransCoder~\cite{roziere2020transcoder} demonstrated unsupervised neural translation among C++, Java, and Python, evaluating functional correctness through unit tests. 
Pan et al.~\cite{pan2024lost} identified frequent functional bugs and a taxonomy of failures in LLM-generated translations of real-world programs. 
FLOURINE~\cite{eniser2024flourine} uses differential fuzzing to compare source programs with LLM-generated Rust translations and feeds counterexamples back to the model for iterative repair. 
AlphaTrans~\cite{ibrahimzada2025alphatrans} scales LLM-driven translation to repository-level programs by combining program analysis, compositional
translation, and multi-level execution-based validation. 
These systems primarily improve translation coverage, generation quality, or validation of individual generated programs.

\sys distills recurring LLM-generated transformations into executable rules to amortize inference cost. Reuse can also propagate translation errors, so we audit the rules' semantic preservation claims over a modeled Fortran-to-Python subset. Execution-based validation provides concrete feedback, while formal reasoning checks whether a rule generalizes beyond the observed instances.

\vspace{5pt}
\noindent \textbf{Verified compilation and verified rewrite rules.}
A large body of work provides strong correctness guarantees by formally verifying compilers or program transformations. 
CompCert~\cite{leroy2009compcert} establishes semantic preservation for a realistic optimizing C compiler using machine-checked proofs. 
Subsequent systems have explored how to make verified compilation more extensible. XCert~\cite{tatlock2010xcert} extends CompCert with a verified execution engine for optimizations written in a domain-specific language, allowing new optimizations to inherit the correctness guarantee of the engine. 
More recently, Pyrosome~\cite{jamner2025pyrosome} develops an extensible framework in which new language features and compilation rules can be added while reusing existing correctness results.

Our work is particularly related to systems that verify individual rewrite rules. 
Alive~\cite{lopes2015alive} provides a domain-specific language for LLVM peephole optimizations and automatically proves rules correct or produces counterexamples, exposing eight incorrect optimizations among more than 300 formalized rules. 
Formal models have similarly been used to show that seemingly standard compiler transformations can be invalid under subtle language semantics, such as the C11 memory model~\cite{vafeiadis2015c11opt}. 
Verified tensor scheduling~\cite{liu2022verifiedtensor} constructs source-to-source optimizations from reusable, semantics-preserving high-level rewrites.

\sys shares this focus on reusable transformations but derives its rules from translation and debugging in an existing LLM-driven system. The rules remain production Python translation logic. For a selected subset, LLM assistance produces mechanized Fortran and Python semantics and Lean preservation obligations. Human experts review both the semantic models and each obligation's correspondence with its executable rule. Lean checks the resulting proofs or refutations. This provides a formal audit of production rules without requiring the translator to be implemented in a verified framework.

\vspace{5pt}
\noindent \textbf{Translation validation and equivalence checking.}
Translation validation checks semantic consistency between source and target programs produced by a particular compilation, rather than proving the compiler implementation correct. 
Necula~\cite{necula2000translation} introduced translation validation for an optimizing compiler by comparing programs before and after optimization passes. 
Later work improves the ability to align and compare programs whose source and target control structures differ. 
Semantic program alignment, for example, uses concrete executions to discover alignments from which equivalence-friendly product programs are constructed \cite{churchill2019alignment}.

Translation validation has also been deployed at substantial scale in production compilers. 
Alive2~\cite{lopes2021alive2} performs bounded translation validation over LLVM IR without requiring modifications to LLVM and has uncovered numerous miscompilations and ambiguities in the IR semantics. 
ARM-TV~\cite{berger2025armtv} extends this approach across the larger semantic gap from LLVM IR to AArch64 code and reports dozens of previously unknown backend miscompilations. 
Related work on trustworthy program verifiers validates the actual translation performed by verifier front-ends by generating independently checkable proofs relating source programs to their generated intermediate-verification-language programs \cite{parthasarathy2024trustworthy}. 
These approaches importantly emphasize that correctness of an idealized translation does not by itself establish correctness of the implementation that performs the translation.

Our audit targets reusable translation rules. A checked proof establishes preservation for every modeled instance satisfying the rule's premises, so the assurance can be reused with the rule. A checked refutation identifies a counterexample to that generalization. Applying either result to production depends on human-reviewed correspondences between language behavior and mechanized semantics and between executable rules and obligations.


\section{Conclusion}
\label{sec:conclusion}

We presented \sys, an approach to bring assurance and determinism for LLM-driven Fortran-to-Python translation.
\sys turns recurring LLM-generated transformations into executable rules for deterministic reuse and introduces Lean proof to provide assurance and an additional feedback channel.
We evaluated \sys on 17 benchmarks from 15 scientific software.
The translated Python achieved bitwise agreement with native Fortran on all evaluated benchmark tests under the recorded execution and initialization configurations.
Across these software, \sys accumulated 193 executable rules and 188 audit records, of which 121 have checked soundness proofs, including 31 established after counterexample-guided refinement.
Cross-software reuse exceeded 90\% for 14 of the 15 systems, and LLM usage costs were 44.2\% lower than direct LLM translation even with additional assurance and determinism.
Further work will expand the modeled language fragments and interoperability support and explore application to other language pairs.

\section*{Data-Availability Statement}
\sys is released as part of RecastEngine~\cite{recastengine2026}.


\clearpage
\bibliographystyle{ACM-Reference-Format}
\bibliography{references,motivation_nint}

\clearpage
\appendix
\begingroup
\setlength{\intextsep}{4pt}
\captionsetup[table]{skip=3pt}
\numberwithin{table}{section}
\numberwithin{figure}{section}
\section{Executable Rule Categories}
\label{app:rule-categories}
\begin{table}[H]
\caption{Categories of the 193 registered executable rule definitions. The registry contains 29 categories, with the 18 smaller categories combined in the final entry.}
\label{tab:registered-rules}
\centering \fontsize{8}{9.4}\selectfont \setlength{\tabcolsep}{4pt} \setlength{\arrayrulewidth}{0.4pt} \arrayrulecolor{black} \renewcommand{\arraystretch}{1.08}
\begin{tabular}{@{}l|c||l|c||l|c||l|c@{}}
\specialrule{0.8pt}{0pt}{0pt}
\multicolumn{1}{c}{\textbf{Category}} & \multicolumn{1}{c}{\textbf{\#}} & \multicolumn{1}{c}{\textbf{Category}} & \multicolumn{1}{c}{\textbf{\#}} & \multicolumn{1}{c}{\textbf{Category}} & \multicolumn{1}{c}{\textbf{\#}} & \multicolumn{1}{c}{\textbf{Category}} & \multicolumn{1}{c}{\textbf{\#}} \\
\specialrule{0.4pt}{0pt}{0pt}
Precision/kind & 23 & Arithmetic & 9 & Control flow & 17 & Indexing & 9 \\

Env. iface. & 17 & Memory ops. & 9 & Procedures & 12 & Reductions & 7 \\

Array trans. & 10 & Spec. intr. & 7 & I/O & 10 & Other (18) & 63 \\
\bottomrule
\end{tabular}
\end{table}

\section{Lean Proof Categories}
\label{app:lean-proof-categories}
\begin{table}[H]
\caption{Families of Lean proofs with representative theorem declarations. The examples include preservation theorems, results under additional premises, and checked refutations.}
\label{tab:lean-proof-categories}
\centering \fontsize{8}{9.4}\selectfont \setlength{\tabcolsep}{4pt} \setlength{\arrayrulewidth}{0.4pt} \arrayrulecolor{black} \renewcommand{\arraystretch}{1.08}
\begin{tabular}{@{}l|l||l|l@{}}
\specialrule{0.8pt}{0pt}{0pt}
\multicolumn{1}{c}{\textbf{Category}} & \multicolumn{1}{c}{\textbf{Representative theorem}} & \multicolumn{1}{c}{\textbf{Category}} & \multicolumn{1}{c}{\textbf{Representative theorem}} \\
\specialrule{0.4pt}{0pt}{0pt}
Literals/names & \begin{tabular}[t]{@{}l@{}}\texttt{FNP\_R0001\_sound}\\\texttt{FNP\_R0010\_sound}\end{tabular} & Array stores & \texttt{FNP\_R0113\_sound} \\
Arithmetic & \texttt{FNP\_R0026\_sound} & Array transforms & \texttt{FNP\_R0106\_cshift\_sound} \\
Comparison/logic & \texttt{FNP\_R0034\_sound} & Reductions & \texttt{FNP\_R0101\_sound} \\
Character ops. & \texttt{FNP\_R0085\_trim\_sound} & Procedure calls & \texttt{FNP\_R0141\_refuted\_unknown} \\
Scalar intrinsics & \texttt{FNP\_R0062\_sound} & Control flow & \texttt{FNP\_R0123\_sound} \\
Elemental lifting & \texttt{lift2\_rel} & Sequence assoc. & \texttt{FNP\_R0056\_sound} \\
Array indexing & \texttt{element\_sound} & Composition & \begin{tabular}[t]{@{}l@{}}\texttt{trans\_sound}\\\texttt{transStmt\_sound}\end{tabular} \\
Array sections & \texttt{FNP\_R0044\_sound} & Whole kernels & \texttt{saxpy\_sound} \\
\bottomrule
\end{tabular}
\end{table}

\section{Single-source Rule Coverage}
\label{app:single-source-rule-reuse}
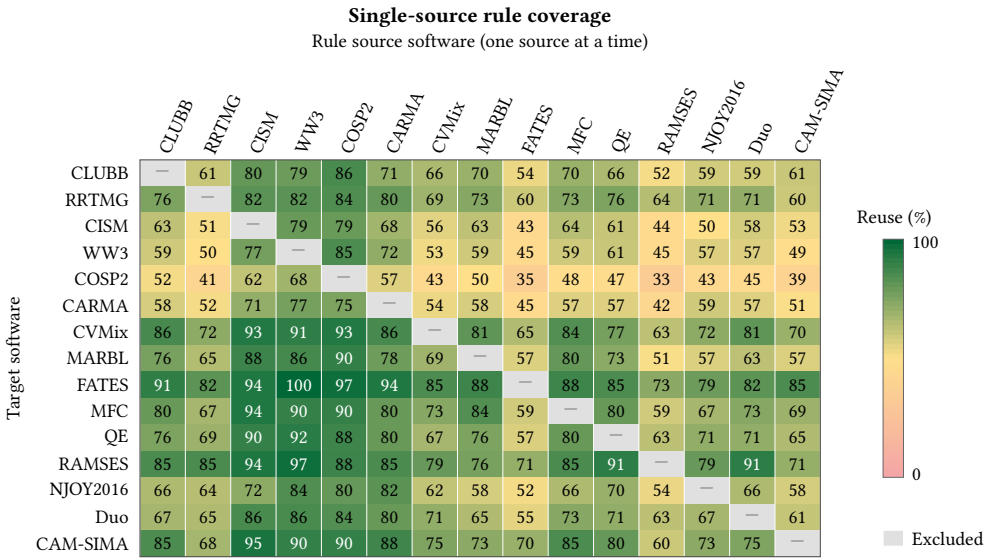
\begin{figure}[!ht]
\centering
\def\rqSingleReuseNumerators{{
  0,34,45,44,48,40,37,39,30,39,37,29,33,33,34, 
  34,0,37,37,38,36,31,33,27,33,34,29,32,32,27, 
  45,37,0,57,57,49,40,45,31,46,44,32,36,42,38, 
  44,37,57,0,63,53,39,44,33,44,45,33,42,42,36, 
  48,38,57,63,0,52,40,46,32,44,43,30,40,41,36, 
  40,36,49,53,52,0,37,40,31,39,39,29,41,39,35, 
  37,31,40,39,40,37,0,35,28,36,33,27,31,35,30, 
  39,33,45,44,46,40,35,0,29,41,37,26,29,32,29, 
  30,27,31,33,32,31,28,29,0,29,28,24,26,27,28, 
  39,33,46,44,44,39,36,41,29,0,39,29,33,36,34, 
  37,34,44,45,43,39,33,37,28,39,0,31,35,35,32, 
  29,29,32,33,30,29,27,26,24,29,31,0,27,31,24, 
  33,32,36,42,40,41,31,29,26,33,35,27,0,33,29, 
  33,32,42,42,41,39,35,32,27,36,35,31,33,0,30, 
  34,27,38,36,36,35,30,29,28,34,32,24,29,30,0 
}}
\def\rqSingleReuseDenominators{{56,45,72,74,92,69,43,51,33,49,49,34,50,49,40}}
\definecolor{rqSingleReuseLow}{HTML}{F4A6A6}
\definecolor{rqSingleReuseMid}{HTML}{FEE08B}
\definecolor{rqSingleReuseHigh}{HTML}{006837}
\def\rqSingleReuseSetColor#1{%
  \ifdim#1pt<50pt
    \pgfmathsetmacro{\rqSingleReuseBlend}{2*#1}%
    \colorlet{rqSingleReuseCellColor}{rqSingleReuseMid!\rqSingleReuseBlend!rqSingleReuseLow}%
  \else
    \pgfmathsetmacro{\rqSingleReuseBlend}{2*(#1-50)}%
    \colorlet{rqSingleReuseCellColor}{rqSingleReuseHigh!\rqSingleReuseBlend!rqSingleReuseMid}%
  \fi
}
\cachetikzfigure{single-source-rule-reuse}{figs/single_source_rule_reuse.tex}
\begin{tikzpicture}[x=0.60cm,y=-0.35cm,
  every node/.style={font=\fontsize{7.2}{8.6}\selectfont,inner sep=1pt}]
  \node[font=\fontsize{8}{9.6}\selectfont\bfseries] at (7.5,-5.40)
    {Single-source rule coverage};
  \node at (7.5,-4.46) {Rule source software (one source at a time)};
  \node[rotate=90] at (-2.75,7.5) {Target software};
  \foreach \software [count=\idx] in
    {CLUBB,RRTMG,CISM,WW3,COSP2,CARMA,CVMix,MARBL,FATES,MFC,QE,RAMSES,NJOY2016,Duo,CAM-SIMA} {
    \node[anchor=east] at (-0.18,\idx-0.5) {\software};
    \node[anchor=west,rotate=65] at (\idx-0.5,-0.17) {\software};
  }
  \foreach \testrow in {1,...,15} {
    \foreach \sourcecol in {1,...,15} {
      \ifnum\testrow=\sourcecol
        \fill[black!12] (\sourcecol-1,\testrow-1) rectangle (\sourcecol,\testrow);
        \node[text=black!55] at (\sourcecol-0.5,\testrow-0.5) {\textemdash};
      \else
        \pgfmathsetmacro{\covered}{\rqSingleReuseNumerators[(\testrow-1)*15+\sourcecol-1]}
        \pgfmathsetmacro{\denominator}{\rqSingleReuseDenominators[\testrow-1]}
        \pgfmathsetmacro{\shade}{100*\covered/\denominator}
        \pgfmathtruncatemacro{\celllabel}{round(\shade)}
        \rqSingleReuseSetColor{\shade}
        \fill[rqSingleReuseCellColor]
          (\sourcecol-1,\testrow-1) rectangle (\sourcecol,\testrow);
        \ifnum\celllabel>88
          \node[text=white] at (\sourcecol-0.5,\testrow-0.5) {\celllabel};
        \else
          \node[text=black] at (\sourcecol-0.5,\testrow-0.5) {\celllabel};
        \fi
      \fi
    }
  }
  \draw[step=1,white,line width=0.3pt] (0,0) grid (15,15);
  \draw[black!70,line width=0.4pt] (0,0) rectangle (15,15);
  \begin{scope}[xshift=0.5cm]
  \node at (15.8,2.2) {Reuse (\%)};
  \foreach \step in {0,...,49} {
    \pgfmathtruncatemacro{\legendshade}{100*\step/49}
    \rqSingleReuseSetColor{\legendshade}
    \fill[rqSingleReuseCellColor]
      (15.55,{12-0.18*\step}) rectangle (16.0,{12-0.18*(\step+1)});
  }
  \draw[black!60,line width=0.3pt] (15.55,3) rectangle (16.0,12);
  \node[anchor=west] at (16.12,3.1) {100};
  \node[anchor=west] at (16.12,11.9) {0};
  \filldraw[fill=black!12,draw=white,line width=0.3pt]
    (15.55,14.0) rectangle (16.0,14.6);
  \node[anchor=west] at (16.12,14.3) {Excluded};
  \end{scope}
\end{tikzpicture}
\caption{Single-source coverage of observed translation rules across 15 software
systems. Each column independently supplies only that source software's observed
rules. Each number is the rounded percentage of the target software's observed
rule types that also occur in the column software:
$100\,|R(t)\cap R(s)|/|R(t)|$. Repeated events count once per rule type.
The denominator belongs to the row target, so coverage is directional.
Gray diagonal cells exclude self-comparisons.}
\Description{A 15 by 15 heatmap of retrospective translation-rule overlap.
Rows identify target software; columns identify independent rule sources.
Numeric labels and a light-red--yellow--green scale show percentage coverage
from zero to one hundred. Gray diagonal cells exclude self-comparisons.}
\label{fig:single-source-rule-reuse}
\end{figure}

\clearpage
\section{Frequently Reused Translation Rules}
\label{app:rule-reuse-frequency}
\begin{table}[H]
\centering
\fontsize{7.2}{8.6}\selectfont
\setlength{\tabcolsep}{1.6pt}
\renewcommand{\arraystretch}{1.08}
\caption{Per-software event counts for the ten most frequently observed
registered translation rules, ordered by total frequency.
Counts include all recorded outcomes, including rewritten, discarded, and
rollback-marked emissions. CAM-SIMA combines its three cases, counting
shared events once. Zero denotes no captured matching event.}
\label{tab:rule-reuse-counts}
\begin{tabular}{@{}l*{15}{r}@{}}
\toprule
\textbf{Rule} & \rotatebox{65}{\textbf{QE}} & \rotatebox{65}{\textbf{MFC}} & \rotatebox{65}{\textbf{NJOY2016}} & \rotatebox{65}{\textbf{RAMSES}} & \rotatebox{65}{\textbf{Duo}} & \rotatebox{65}{\textbf{CVMix}} & \rotatebox{65}{\textbf{MARBL}} & \rotatebox{65}{\textbf{RRTMG}} & \rotatebox{65}{\textbf{COSP2}} & \rotatebox{65}{\textbf{CARMA}} & \rotatebox{65}{\textbf{FATES}} & \rotatebox{65}{\textbf{CISM}} & \rotatebox{65}{\textbf{WW3}} & \rotatebox{65}{\textbf{CAM-SIMA}} & \rotatebox{65}{\textbf{CLUBB}} \\
\midrule
\textrm{R-name.reference} & 14295 & 18215 & 5792 & 3609 & 900 & 2239 & 7033 & 6769 & 11629 & 12014 & 1620 & 21853 & 11826 & 1513 & 32165 \\
\textrm{R-index.shift\_unit} & 1929 & 3631 & 1094 & 1172 & 197 & 430 & 3201 & 3825 & 3831 & 5424 & 46 & 7735 & 3625 & 935 & 12022 \\
\shortstack[l]{\textrm{R-arithmetic.}\\\textrm{scalar\_fallback}} & 9402 & 8057 & 2594 & 1004 & 450 & 854 & 1988 & 2877 & 798 & 4303 & 477 & 4019 & 1952 & 581 & 4689 \\
\shortstack[l]{\textrm{R-index.}\\\textrm{array\_reference}} & 2281 & 6853 & 990 & 807 & 141 & 648 & 1852 & 3780 & 2875 & 1439 & 72 & 4830 & 4348 & 595 & 6822 \\
\shortstack[l]{\textrm{R-literal.}\\\textrm{hoisted\_reference}} & 7218 & 884 & 1718 & 66 & 97 & 128 & 351 & 18958 & 1822 & 1783 & 134 & 1371 & 1191 & 47 & 435 \\
\shortstack[l]{\textrm{R-assignment.}\\\textrm{ordinary}} & 3956 & 2807 & 1791 & 564 & 257 & 545 & 1173 & 2926 & 3170 & 3178 & 237 & 2989 & 3991 & 343 & 2826 \\
\shortstack[l]{\textrm{R-literal.}\\\textrm{integer\_inline}} & 834 & 3393 & 1426 & 536 & 454 & 306 & 270 & 2125 & 2074 & 1304 & 16 & 2685 & 2127 & 132 & 2981 \\
\textrm{R-component.path} & 0 & 2575 & 0 & 4 & 88 & 223 & 2493 & 82 & 4073 & 3954 & 55 & 3482 & 422 & 9 & 2825 \\
\shortstack[l]{\textrm{R-index.}\\\textrm{slice\_ascending}} & 229 & 202 & 0 & 42 & 10 & 65 & 1963 & 450 & 6588 & 1246 & 6 & 3575 & 157 & 99 & 3821 \\
\shortstack[l]{\textrm{R-index.}\\\textrm{shift\_declared}} & 177 & 7484 & 0 & 409 & 2 & 0 & 29 & 821 & 63 & 79 & 0 & 656 & 1330 & 0 & 4 \\
\bottomrule
\end{tabular}
\end{table}

\endgroup

\end{document}